\documentclass[journal]{IEEEtran}
\ifCLASSINFOpdf
   \usepackage[pdftex]{graphicx}
   \graphicspath{{Figure/}{../jpeg/}}
   \DeclareGraphicsExtensions{.pdf,.jpeg,.png}
\else
   \usepackage[dvips]{graphicx}
   \graphicspath{{../eps/}}
   \DeclareGraphicsExtensions{.eps}
\fi
\usepackage{booktabs} 

\usepackage[colorlinks, citecolor=black, urlcolor=blue]{hyperref}
\usepackage{makecell}

\usepackage{amsmath}
 
\usepackage{cite}

\usepackage{graphicx}

\usepackage{multirow}

\usepackage{enumerate}

\usepackage{float}

\usepackage{algorithm}
\usepackage{algorithmic}

\makeatletter
\newcommand{\removelatexerror}{\let\@latex@error\@gobble}
\makeatother

\ifCLASSOPTIONcompsoc
  \usepackage[caption=false,font=normalsize,labelfont=sf,textfont=sf]{subfig}
\else
  \usepackage[caption=false,font=footnotesize]{subfig}
\fi
\begin{document}
%
\title{Unsignalized Intersection Management Strategy\\ for Mixed Autonomy Traffic Streams}

%
%
%

\author{Junjie~Zhou,~Zhaokun~Shen,~Xiaofan~Wang,
        and~Lin~Wang
\thanks{This work was supported by the National Key R$\&$D Program of China
with No. 2023YFB4706802, and the National Natural Science Foundation of
China with No. 62373245 and 62336005, and in part by the “Dawn” Program of Shanghai
Ed. \emph{(Corresponding author: Lin Wang.)}}
\thanks{Junjie Zhou, Zhaokun Shen and Lin Wang are with the Department of Automation, Shanghai Jiao Tong University, and Key Laboratory of System Control and Information Processing, Ministry of Education of
China, Shanghai 200240, China (e-mail: junjie\_zhou@sjtu.edu.cn; szk951124@sjtu.edu.cn; wanglin@sjtu.edu.cn).}
\thanks{Xiaofan Wang is with the Department of Automation, Shanghai Jiao
Tong University, the School of Mechatronic Engineering and Automation, Shanghai University, and the School of Electrical and Electronic
Engineering, Shanghai Institute of Technology, Shanghai, 201418, China (e-mail: xfwang@sjtu.edu.cn).
}
}

%
%

\markboth{PREPRINT under review}%
{Shell \MakeLowercase{\textit{et al.}}: Bare Demo of IEEEtran.cls for IEEE Journals}
%



\maketitle

\begin{abstract}
With the rapid development of connected and automated vehicles (CAVs) and intelligent transportation infrastructure, CAVs, connected human-driven vehicles (CHVs), and un-connected human-driven vehicles (HVs) will coexist on the roads in the future for a long time. This paper comprehensively considers the different traffic characteristics of CHVs, CAVs, and HVs, and systemically investigates the unsignalized intersection management strategy from the upper decision-making level to the lower execution level. The unsignalized intersection management strategy consists of two parts: the heuristic priority queues based right of way allocation (HPQ) algorithm and the vehicle planning and control algorithm. In the HPQ algorithm, a vehicle priority management model considering the difference between CAVs, CHVs, and HVs, is built to design the right of way management for different types of vehicles. In the lower level for vehicle planning and control algorithm, different control modes of CAVs are designed according to the upper-level decision made by the HPQ algorithm. Moreover, the vehicle control execution is realized by the model predictive controller combined with the geographical environment constraints and the unsignalized intersection management strategy. The proposed strategy is evaluated by simulations, which show that the proposed intersection management strategy can effectively reduce travel time and improve traffic efficiency. Results show that the proposed method can decrease the average travel time by 5\% to 65\% for different traffic flows compared with the comparative methods. The intersection management strategy captures the real-world balance between efficiency and safety for future intelligent traffic systems. 
\end{abstract}


%
\IEEEpeerreviewmaketitle



\section{INTRODUCTION}
%
%
%
%
\IEEEPARstart{I}{n recent} years, traffic congestion has become a critical social problem that troubles people's lives and hinders social development \cite{8950042}. With the development of autonomous driving technology and the maturity of communication technology, connected and automated vehicles (CAVs) have gradually come into the public horizon. It is the primary direction for researchers to improve intersection efficiency through vehicle coordination under safe conditions. Traffic signal control methods have good effects on intersections with relatively large traffic volumes and strong traffic capacity \cite{5978226}. But for intersections with small traffic volumes and weak traffic capacity, signal control will cause unnecessary travel time loss. 

Unsignalized intersections have been considered an effective solution to improve traffic efficiency in the past few years \cite{9403993}. While unsignalized intersections provide effective tools to reduce energy consumption and traffic delays, it is also necessary to improve safety to reduce the number of accidents \cite{8479331, 9261970}. As the penetration rate of connected vehicles is increasing rapidly, it
is expected that such systems will become more practical in the near future. Besides, according to the study in \cite{bifulco2022decentralized, ma2021trajectory}, governments worldwide are increasingly promoting the deployment of Vehicle-to-Everything (V2X) technologies.
Considering the traffic environment of mixed connected human-driven vehicles (CHVs), un-connected human-driven vehicles (HVs), and CAVs will exist for an extended period of time in the future, efficient and reasonable intersection management strategies should be developed to ensure vehicle safety and improve the traffic efficiency for mixed autonomy traffic streams. 

This paper focuses on the design of unsignalized intersections management strategy for mixed autonomy traffic streams based on the vehicle-to-infrastructure (V2I) architecture, which consists of the heuristic priority queues based right of way allocation (HPQ) algorithm and vehicle planning and control algorithm for the mixed traffic flow intersections. 
The HPQ algorithm is used to manage the right of way of vehicles at intersections to maximize the utilization efficiency of the conflict area. Through the geographic modeling of the intersection, the trajectories and conflict table of vehicles passing through the intersection are formulated, and the traffic condition constraints are used to ensure the safety of vehicles. According to the decision made by the HPQ algorithm at the intersection, the vehicle planning and control algorithm is designed based on four control modes, including car following mode, cruise mode, waiting mode, and conflict-solving mode. 

Furthermore, the improved model predictive control is applied to realize the four control modes.
The main contributions of this paper are as follows:
\begin{itemize}
\item We propose an integrated heuristic priority queues based intersection management strategy spanning from upper decision-making levels to lower execution levels. Distinguished from previous research, this paper proposes an enhanced HPQ algorithm tailored to address trajectory conflicts and prioritize management conditions for CAVs, CHVs, and HVs. The overall time complexity of our algorithm is $O(n)$, ensuring real-time coordination and control of unsignalized intersections while reducing hardware computing device overhead.
\item Four vehicle control modes are proposed based on the developed HPQ algorithm, with model predictive control facilitating smooth transitions between different control modes. Our robust scheduling and control strategy fully account for the characteristics of CAVs, CHVs, and HVs, with CAVs proactively avoiding trajectory conflicts with HVs to ensure safe and efficient traffic flow.
\item To validate the safety and effectiveness of our proposed method, we conducted a hybrid simulation using SUMO and PreScan, evaluating both macroscopic and microscopic aspects. Our developed unsignalized intersection management strategy outperforms three existing state-of-the-art methods in terms of travel time and the number of halts. Furthermore, our method is applicable to intersections of various sizes and types. To the best of our knowledge, this is the first time that an unsignalized intersection management method can efficiently coordinate mixed traffic flows, containing CAVs, CHVs, and HVs simultaneously.
\end{itemize}


The rest of the paper is organized as follows. Section II reviews the related works. Section III introduces the preliminary and system architecture of the proposed strategy. Section IV details the design of the HPQ algorithm, including the priority management model and right of way management model. Section V details the design of vehicle planning and control algorithm, including the path planning model, speed planning model and modified model predictive control for vehicle control modes. The simulation and performance comparison are presented in Section VI. While section VII concludes the paper.


\begin{table*}[!ht]

\renewcommand{\arraystretch}{1.5}
{\caption{SUMMARY OF UNSIGNALIZED INTERSECTION MANAGEMENT METHODS}}

\label{table_4}
\centering

\resizebox{2\columnwidth}{!}
{
\begin{tabular}{ccccccccc}
\toprule
\toprule

\multirow{2}{*}{Category} &  \multirow{2}{*}{Study} & \multirow{2}{*}{Method} & \multirow{2}{*}{Intersection Type}  & \multirow{2}{*}{Control Policy} & \multicolumn{3}{c}{ Mixed Traffic Flow} & \multirow{2}{*}{\makecell[c]{Validation Methods}} \\
\cline{6-8}
&&&&& CAV  & CHV & HV\\
\hline
\multirow{11}{*}{\makecell[c]{Centralized System}}   &  \cite{deng2020conflict} & Conflict Duration Graph & Signal-Free Intersection & Reservation Policy & *  &  &  & MATLAB  \\
&  \cite{lin2019graph} & Cycle Removal Algorithm & One-lane Intersection & Cycle Removal & *  &   &  &  Numerical Test \\
&  \cite{choi2019reservation} & Reservation-based Scheduling & Unsignalized Intersection & Reservation Policy &  *  &   &  &  Macro-simulation \\
&\cite{10403872} &  Monte Carlo Tree Search & \makecell[c]{One-lane\\ Unsignalized Intersection}  & \makecell[c]{Learning-based\\ Iterative Optimization}  & *   &  &  & SUMO\\
& \cite{9583858} & \makecell[c]{Transfer Deep\\ Reinforcement Learning } & \makecell[c]{One-lane\\ Unsignalized Intersection}  &Priority Policy & *  &  &  & OpenAI gym \\
& \cite{9583896} & Deep Reinforcement Learning & Three-way Intersection & Priority Policy & *  &   & * & SUMO  \\
& \cite{9204585} & Reinforcement Learning  & \makecell[c]{One-lane\\ Unsignalized Intersection} & Coordination Policy & *  &  &  & Numerical Simulation \\
& \cite{10186800} &Reinforcement Learning & \makecell[c]{One-lane\\ Unsignalized Intersection}  & Priority Policy & *  &   &  *  &  Highway-env \\
& \cite{8848796} & Absolute Value Programming & Unsignalized Intersection & Priority Policy & *  &   &    & SUMO \\
&  \cite{10452809} & \makecell[c]{Reservation-prioritization-based\\ Cooperative Control} & Unsignalized Intersection & Reservation Policy &  * &  & *  & SUMO \\
& \cite{9827300}  &  Cooperative Maneuver Planning  &  Unsignalized Intersection   &  Reservation Policy  &  *   &   *  &  &   SUMO \\
&  \cite{10224258}  &  \makecell[c]{Learning-based\\ Iterative Optimization}  &  Unsignalized Intersection   &  \makecell[c]{Cluster-based\\ Motion Planning}  &  *   &   &  &  SUMO  \\
&   \cite{10360227}  &   Conflict Graph Tree Search  &  \makecell[c]{Multi-lane\\ Unsignalized Intersection}   &  Priority Policy  &  *    &   &  &  Python \\
\hline
\multirow{6}{*}{ \makecell[c]{Decentralized System}}    &   \cite{PENG2021100017}  &  Deep Reinforcement Learning  & Double-lane Intersection  &  Platoon Control  &    *  &  &  *   &  SUMO Simulation  \\
&  \cite{bifulco2022decentralized}  &  \makecell[c]{Fully-distributed  \\ Control Protocol}  &  \makecell[c]{One-lane\\ Unsignalized Intersection} &   Priority Policy  &  *  &  *   &  &  MiTraS  \\
&  \cite{10021253}  &  \makecell[c]{Vehicle-Platoon-Aware \\ Bi-Level Optimization}  &  \makecell[c]{One-lane\\ Unsignalized Intersection}  &  Reservation Policy  &  *   &   &  *  &  MATLAB \\
&   \cite{ZHOU2022103610}  &  Virtual Platooning    &  Signal-free Intersection  &  \makecell[c]{Adaptive Sliding \\Mode Controller  }  &   *   &  &  &  SUMO \\
&  \cite{9336022}  &  \makecell[c]{Multiagent-based Deep \\Reinforcement Learning}  &  \makecell[c]{Multiple\\ Unsignalized Intersections}  &  Priority Policy  &   *   &   &   &   Python  \\
\hline
\multirow{8}{*}{ \makecell[c]{Distributed System}}    &   \cite{XU2018322} &  Vehicle Conflict Directed Graph  &  \makecell[c]{One-lane\\ Unsignalized Intersection}  &  Priority Policy  &  *   &  &   &  Numerical Test  \\
&  \cite{9800916}  &  \makecell[c]{Depth-First Spanning Tree} &  Unsignalized Intersection &   Reservation Policy  &  *  &   &  &   SUMO Simulation \\
&  \cite{9564753} &  \makecell[c]{Depth-First Spanning Tree} &  \makecell[c]{Three-lane\\ Unsignalized Intersection}  &   Reservation Policy &  * &   &  &   SUMO Simulation \\  
&  \cite{ma2021development} &  Interspersed Traffic Organization   &  Unsignalized Intersection  &  Priority Policy  &  *   &   &  &  VISSIM  \\
&  \cite{4012536}  &  Tree Generation Algorithm  &  \makecell[c]{One-lane\\ Unsignalized Intersection}   &  Priority Policy  &  *   &   &  &  Numerical Simulation \\
&  \cite{8941315} &  Task-area Partition  & \makecell[c]{One-lane\\ Unsignalized Intersection}  &  Priority Policy  &  *  &   &  &  Numerical Simulation \\
& \cite{10440183} &   
\makecell[c]{Distributed Robust\\ Differential Game}   &  Small Unsignalized Intersection  &  Game Theory  &  *    &   &  &  Simulation \\
&  \cite{9740423}  &  Iterative Grouping Algorithm  &  \makecell[c]{Multi-lane\\ Unsignalized Intersection}  &  Formation Control &  *    &   &  &  SUMO \\
\bottomrule
\bottomrule
\end{tabular}
}
\end{table*}

\section{LITERATURE REVIEW}

\subsection{Unsignalized Autonomous Intersection Management}
As summarized in \cite{khayatian2020survey}, there are two approaches to calculating the possible conflict between two vehicles at the intersection, including the expected vehicle trajectories and the spatiotemporal occupancy map. 
The system architecture of unsignalized intersections can be broadly classified into centralized system, decentralized system, and distributed system. The relevant
works on the unsignalized intersection management methods
are summarized in Table I.

At the centralized control intersection, vehicles obtain traffic schemes from the Intersection Control Unit (ICU) through V2I communications.
The autonomous intersection management (AIM) of the centralized system follows the server-client scheme, which can be divided into reservation-based and priority-based intersection management. The conflict duration graph is proposed in \cite{deng2020conflict} to resolve collisions and scheduling problems for CAVs at unsignalized intersections. 
Petri net and cycle removal algorithms are used in \cite{lin2019graph} to model intersecting scheduling problems. The tree search algorithm is utilized in \cite{choi2019reservation, 10403872} for optimal passing order calculation.
Besides, reinforcement learning algorithms are also widely used in \cite{9583858, 9583896, 9204585, 10186800} for centralized intersection management.
Centralized methods are computationally intensive and time-consuming, especially for large intersections or high traffic volumes. Consequently, they place significant demands on computing resources and algorithmic complexity.

Decentralized intersection management methods aim to mitigate the computational burden of centralized controllers by distributing tasks across multiple shared computing nodes. In these methods, vehicles near the front of the queue often serve as decentralized hubs, forming queues and communicating with the intersection manager to request permission to enter the intersection \cite{bifulco2022decentralized, 10021253, ZHOU2022103610}. 
Platoon formation is widely used in unsignalized autonomous intersection management. CAVs form platoons through V2V communication, and each platoon has a leading vehicle \cite{jin2013platoon,bashiri2018paim}. 
The Vehicle-Platoon-Aware Bi-Level Optimization Algorithm for Autonomous Intersection Management (VPA-AIM) algorithm is proposed in \cite{10021253} to determine the platoon formation scheme of vehicles during the control zone and allocate the passing sequence and timeslots of platoons through the merging area. The VPA-AIM algorithm can be extended to mixed traffic
scenarios. However, this method requires all platoon leaders to be CAVs. It is not feasible if the penetration rates of CAVs are not sufficiently high.

Distributed intersection management methods allocate computational nodes to individual vehicles,
which then collaborate to determine the optimal passing order. 
The distributed robust differential game method is proposed in \cite{10440183} to coordinate driving strategies of CAVs
passing through unsignalized intersections.
The depth-first spanning tree is utilized in \cite{XU2018322} to calculate the right of way of vehicles at the intersection.
Besides, the graph-based coordination method proposed in \cite{9800916, 9564753} enables vehicles to pass through intersections efficiently. It is an effective method for coordinating unsignalized intersections. However, this approach necessitates that all vehicles within the intersection area be equipped with communication and computation devices to facilitate real-time information exchange and conflict resolution. As a
result, distributed methods face challenges when applied to scenarios featuring mixed traffic streams comprising both connected vehicles and un-connected vehicles.

There are relatively few studies on unsignalized control methods for mixed traffic flows where CAVs, CHVs, and HVs coexist.
The control of traditional HVs at intersections poses challenges due to the absence of advanced communication equipment, which renders them unable to respond to directives from intersection control units or receive messages from other vehicles. This limitation complicates multi-vehicle control scenarios at intersections without traffic lights. Certain studies simplify HV behaviors at intersections by assuming strict following to the preceding vehicle. While this simplification aids in the mathematical modeling of multi-vehicle control issues at intersections, it fails to capture real-world scenarios where HVs navigate intersections freely without leading vehicles.
In this paper, we comprehensively consider the different traffic characteristics of CAVs, CHVs, and HVs, and propose a heuristic priority queues based intersection management strategy that can be used for mixed autonomy traffic streams.

\subsection{Mixed Autonomy Traffic Streams Management}
According to the prediction, CAVs will not be fully popularized before 2045 \cite{bansal2017forecasting}. For a long period of time in the future, the traffic environment of mixed CHVs, HVs, and CAVs will always exist. Therefore, it is necessary to study AIM in the mixed autonomy traffic streams. Stone \textit{et al.} first proposed the AIM cooperative driving strategy for mixed traffic streams, which adopts the strategy called First Come First Served (FCFS) manner in the signalized intersection \cite{dresner2008multiagent, 2007Sharing}. HVs obey traffic signals, while CAVs adopt the request-based method to request the right of way at the intersection. In \cite{2017A}, a Hybrid Autonomous Intersection Management (H-AIM) protocol was proposed to grant reservation for vehicles at intersections in FCFS order, which is demonstrated to be effective even in complex traffic scenarios. 

Furthermore, to improve traffic throughput and reduce energy consumption, \cite{lin2017autonomous} established a connected vehicle center (CVC) to control the movement of all vehicles and detect the position and trajectory of HVs through sensors. When the HVs arrive near the stop line, the traffic lights are used to instruct the right of way for these HVs, while the CVC controls the AVs. To reduce the energy consumption of the vehicles at signalized intersections for mixed traffic streams, \cite{zhao2018platoon} proposed a dynamic platoon splitting and merging method to reduce the fuel consumption of the fleet. In addition, a distributed intersection protocol for mixed traffic streams is proposed in \cite{8864572}, where the AVs switch to the synchronous cooperation mode and pass through the intersection without stopping in the case of no HVs entering. 

These existing methods focus on the AIM for mixed autonomy traffic streams at signalized intersections and put forward effective intersection management strategies to improve safety and efficiency. To the best of our knowledge, few studies have considered the difference between CAVs, CHVs, and HVs at unsignalized intersections. In this paper, we formally characterize the differences between CAVs, CHVs, and HVs in handling trajectory conflicts and priority management conditions, and design an improved HPQ algorithm to ensure the safety and travel efficiency of vehicles at unsignalized intersections, which derives a futuristic and practical unsignalized intersection management strategy.

\begin{figure}[!t]
\centering
\includegraphics[width=3.5in]{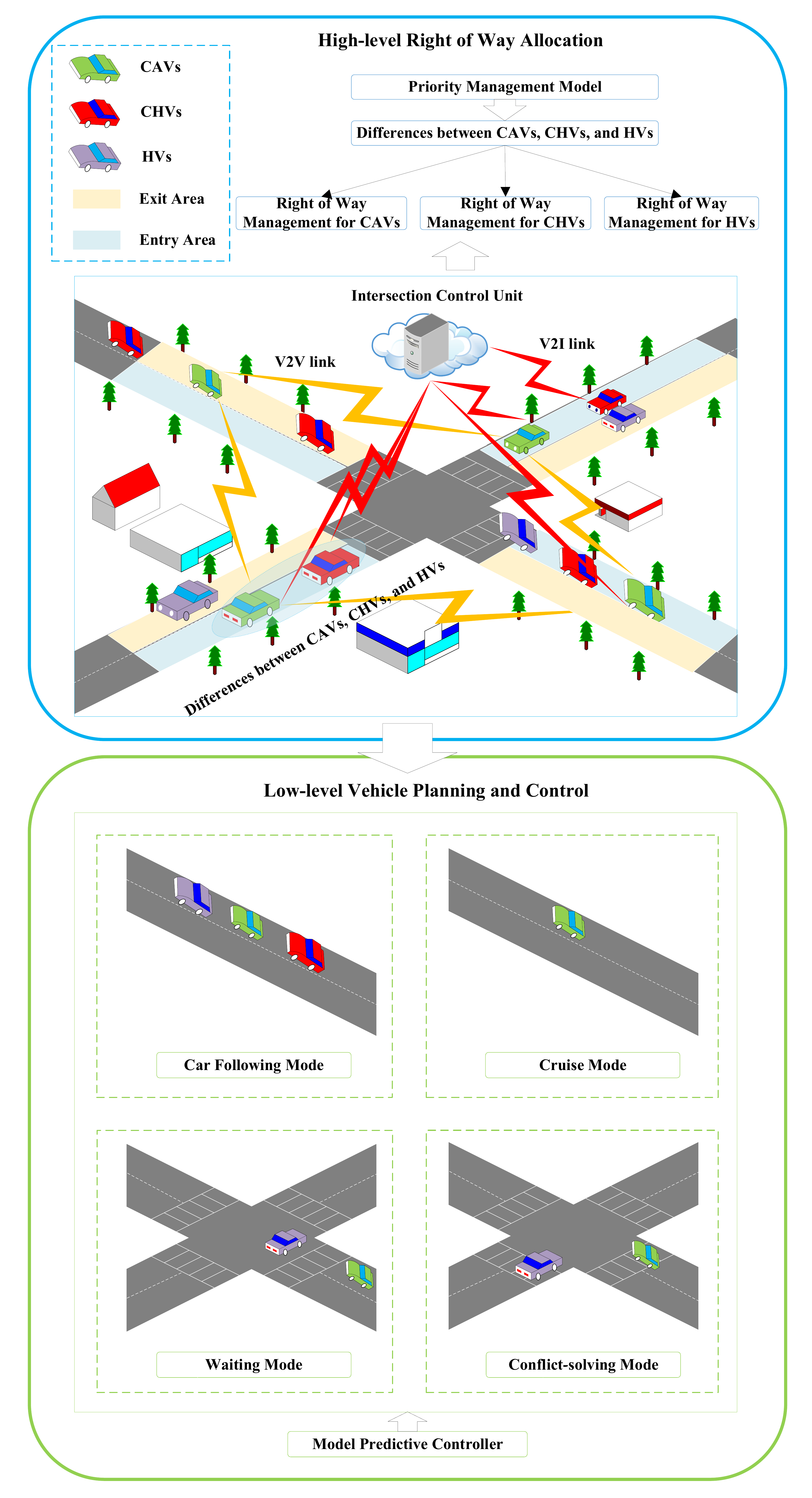}
\caption{Architecture of unsignalized intersection with mixed CAVs, CHVs, and HVs. The system architecture is mainly composed of two levels. The high level is the HPQ algorithm that assigns the right of way to vehicles at the intersection. The low level consists of vehicle planning and control algorithms for CAVs to safely and efficiently resolve potential trajectory conflicts with HVs.}
\label{fig_System structure}
\end{figure}

\section{DESIGN OF SYSTEM ARCHITECTURE}
In this section, a new type of unsignalized AIM system architecture is designed to enable vehicles to pass through the intersection more safely and efficiently. The Cooperative Vehicle Infrastructure System (CVIS) provides hardware and communication support for the intersection. All vehicles at the intersection are assumed to be equipped with an On-Board Unit (OBU) and can transmit the information of the ego vehicle to other vehicles and ICU at the intersection via V2V and V2I communications. Sensor equipment is installed on the roadside to ensure that all messages, such as the speed and position of all vehicles, can be obtained at the intersection. At the same time, it ignores the delay, packet loss, and information attacks in the communication process. When CAVs and CHVs enter the intersection, they will send the estimated driving route to the ICU, and the drivers of CHVs will follow the driving route. A human-machine interface will be installed in the CHVs to display the traffic instructions of the ICU. 

Fig. 1 depicts the architecture of unsignalized intersections with a mix of CAVs, CHVs, and HVs. 
The overall architecture is mainly composed of two levels. The high level is the HPQ algorithm, which is designed in detail in Section IV. It determines the priority management of vehicles and formulates corresponding right of way management methods for the two types of vehicles. The low level includes vehicle planning and control algorithms for CAVs, which are developed in detail in Section V. 

Considering the feasibility and practicability of edging computing, the low-level vehicle planning and control algorithm uses the ICU to collect surrounding vehicle information, which can better ensure the safety of vehicles at the intersection and improve traffic efficiency. Since CAVs have high controllability and intelligence, four control modes, including car following mode, cruise mode, waiting mode, and conflict-solving mode, are designed for CAVs, according to the decision information from the high-level HPQ right of way allocation algorithm. Moreover, the model predictive controller is designed to realize the four control modes of the CAVs. In addition, through the geographic modeling of the intersection, the trajectories and conflict tables of vehicles passing through the intersection are provided in real time.

\begin{figure}[!t]
\centering
\includegraphics[width=3.5in]{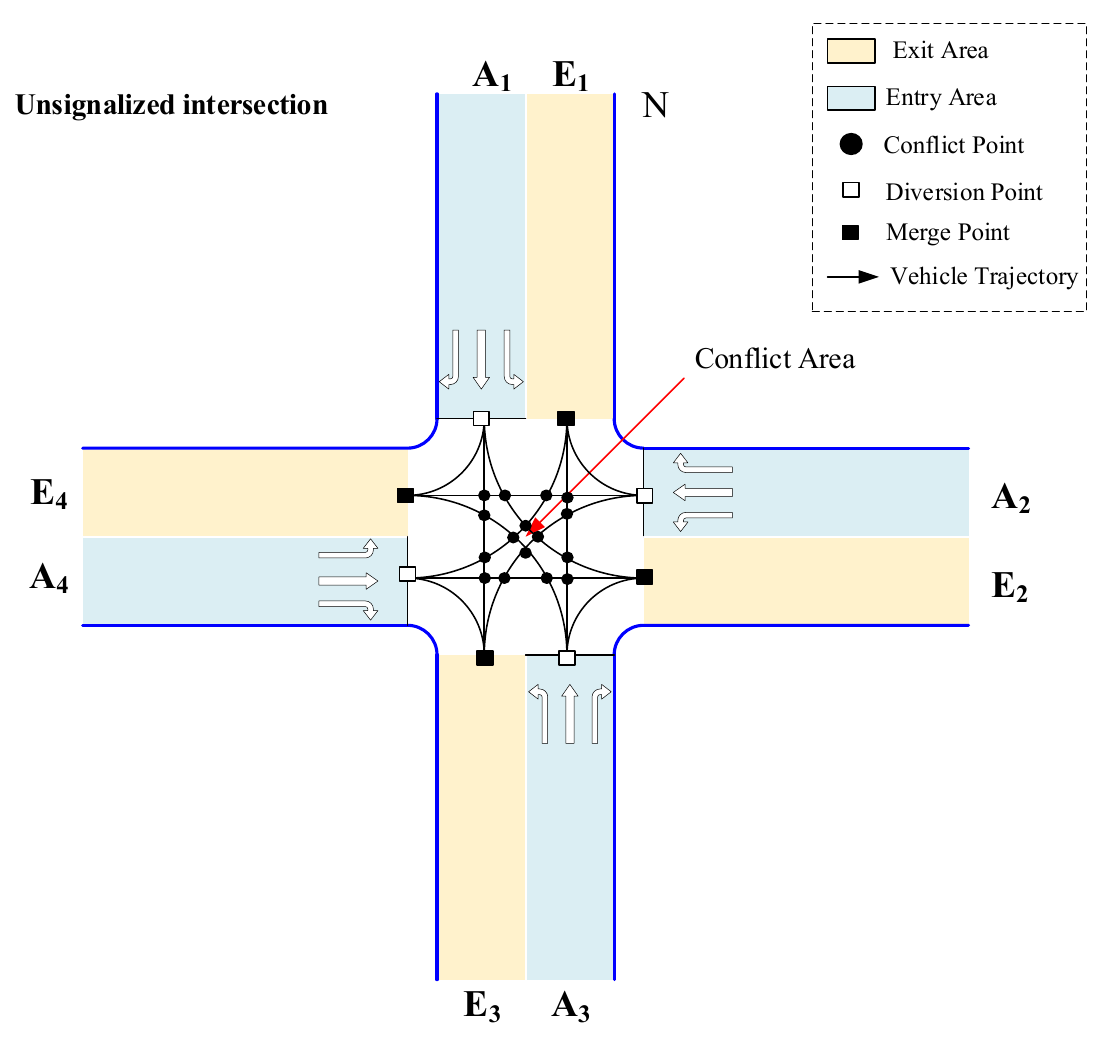}
\caption{Layout of the typical unsignalized intersection.}
\label{fig_sim}
\end{figure}

\begin{table}[!t]
\renewcommand{\arraystretch}{1.3}
\caption{DEFINITION OF SETS, DATA STRUCTURE, VARIABLES AND PARAMETERS}
\label{table_example}
\centering
\begin{tabular}{ll}
\toprule
\toprule
\multicolumn{2}{l}{Sets:}\\
\hline
$S$ & set of all vehicles at the intersection\\
$S_1$ & set of vehicles that have passed through the conflicting area\\
$S_2$ & set of vehicles that have not passed through the conflicting\\& area but have been granted the right of way\\
$S_3$ & set of vehicles that have not obtained the right of way\\
$T_r(V)$ & set of trajectories of vehicle $V$\\

\hline
\multicolumn{2}{l}{Data structure:}\\
\hline
$U$ & struct of the vehicle information containing the id, type,\\& trajectory and other related information\\
$Q_1$ & a priority queue composed of vehicles entering the area\\& at $A_1$ and belonging to the set $S_3$, sorted by priority\\& from high to low\\
$Q_2$ & a priority queue composed of vehicles entering the area\\& at $A_2$ and belonging to the set $S_3$, sorted by priority\\& from high to low\\
$Q_3$ & a priority queue composed of vehicles entering the area\\& at $A_3$ and belonging to the set $S_3$, sorted by priority\\& from high to low\\
$Q_4$ & a priority queue composed of vehicles entering the area\\& at $A_4$ and belonging to the set $S_3$, sorted by priority\\& from high to low\\
$Q_p$ & a priority queue composed of $Q_1[0]$, $Q_2[0]$, $Q_3[0]$, $Q_4[0]$\\&
that are vehicles corresponding to the top elements of\\& $Q_1$, $Q_2$, $Q_3$, $Q_4$\\

\hline
\multicolumn{2}{l}{Variables:}\\
\hline
$I_p$ & the ID of the vehicle that obtains the right of way in the \\& control cycle of the ICU \\
$I_c$ & the ID of the vehicle that has trajectory conflicts with \\& vehicle $V$ \\
$V$ & the ego vehicle at the intersection \\
$\widetilde{V}$ & the vehicles other than $V$ at the intersection  \\
\hline
\multicolumn{2}{l}{Parameters:}\\
\hline
$h$ & the minimum time headway  \\
$T_f$ & the moment the preceding vehicle enters the intersection \\
$T_b$ & the moment the ego vehicle enters the intersection \\
$t_f$ & the travel time of the preceding vehicle from the stop\\& line to the conflict point P  \\
$t_b$ & the travel time of the ego vehicle from the stop\\& line to the conflict point P  \\
$s_f^d$ & the distance of the preceding vehicle from the stop line\\& to the conflict point P   \\
$s_b^d$ & the distance of the ego vehicle from the stop line\\& to the conflict point P   \\
$v_f$ & the speed of the preceding vehicle   \\
$v_b$ & the speed of the ego vehicle   \\
$N_1$ & the number of trajectory conflicts between the vehicle $V$ \\& and other vehicles that have been granted the right of way   \\
$N_2$ & the number of trajectory conflicts between the vehicle $V$ \\& and other vehicles that have not obtained the right of way \\

\bottomrule
\bottomrule
\end{tabular}
\end{table}


\section{HPQ RIGHT OF WAY ALLOCATION ALGORITHM}
In this paper, we consider a typical four-way intersection, as illustrated in Fig. 2. This intersection has four directions, and each direction contains an entry lane and an exit lane. Let $M=\{M_j | j=1,2,3,4\}$ be the set of all directions.
In unsignalized intersections, the right of way allocation algorithm plays a vital role in ensuring safety. A typical conflict scenario is illustrated in Fig. 3, where $P$ is the point of conflict between the two vehicles. Let $T_b$ and $T_f$ be the moments when the ego vehicle and the preceding vehicle enter the intersection, respectively. The essential requirement to ensure the ego vehicle safely pass through the intersection is that: 
\begin{equation}
T_b \ge T_f + h + t_f - t_b, 
\end{equation}
where $t_f = \frac{s_f^d}{v_f}$ and $t_b = \frac{s_b^d}{v_b}$ are the travel time of the preceding vehicle and the ego vehicle from the stop line to the conflict point P, respectively; $s_f^d$ and $s_b^d$ are the distance of the preceding vehicle and the ego vehicle from the stop line to the conflict point P, respectively; $v_f$ and $v_b$ are the speed of the preceding vehicle and the ego vehicle, respectively; $h$ is the minimum time headway, as illustrated in Table II.


For CAVs, the vehicle priority, the location of the conflict point, and the position and speed of the preceding vehicle can be obtained through V2V and V2I communication, and the conflict with the preceding vehicle can be easily handled according to constraint (1). The typical conflict scenario is illustrated in Fig. 3.




\begin{figure}[!t]
\centering
\includegraphics[width=3.3in]{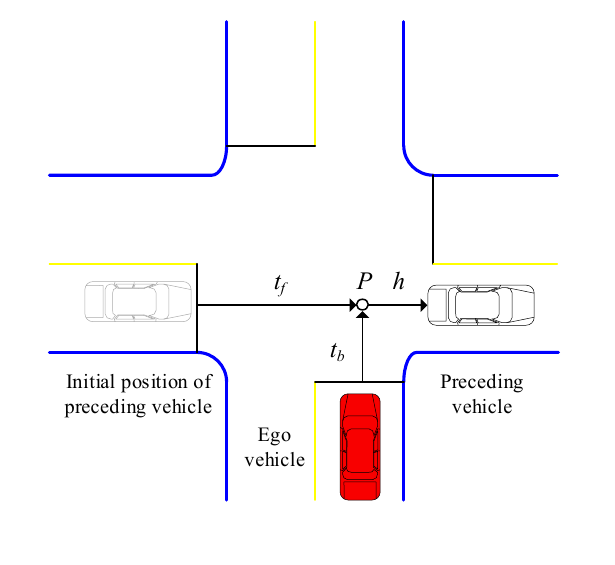}
\caption{A typical conflict scenario.}
\label{fig_Typical_conflict_scenario}
\end{figure}

\subsection{Priority Management Model}
The priority of vehicles at intersections is designed based on the FCFS policy, and the vehicles entering the area of the intersection first will get a lower priority value $P(V)$. The smaller the priority value $P(V)$ obtained by the vehicle, the higher the priority of the vehicle. The priority value $P(V)$ of the vehicle entering the intersection area first is one, and the following vehicles are incremented by one according to the order in which they enter the intersection. When a vehicle obtains the right of way, the priority value $P(V)$ of all vehicles is reduced by one. When the vehicle enters the entry area, the moment of the current vehicle will be recorded by the ICU. When two vehicles have trajectory conflict, the right of way of the vehicles is determined by comparing the size of the time stamp. 

The length of the entry area will influence the driving comfort. In order to guarantee the vehicle can slow down and stop in front of the stop line gracefully. Here, as illustrated in Fig. 2, the length of the entry area $A_l$ satisfies the constraint (2) as follow:
\begin{equation}
A_l \geq \left| \frac{v_{limit}^2}{2a_d} \right|.
\end{equation}
where $v_{limit}$ is the speed limit for the road, and $a_d$ is the maximum acceleration without loss of comfort.


Usually, the priority is not allowed to be changed since changing the priority will easily cause deadlock at the intersection. The deadlock at an intersection means that multiple vehicles fall into a stalemate of cyclical waiting due to the fight for the right of way, and the vehicles can not pass normally at the intersection. 


However, for abnormal situations such as breakdowns in conflict areas, the priority of the vehicle can be changed in the priority management model. If the abnormal vehicle keeps occupying the right of way, it will lead to the deadlock of the vehicles whose priority is lower than its and whose trajectory conflicts with it. In this instance, the ICU will kick the abnormal vehicle out of the priority sorting queue and adjust the priority of the vehicles in the same lane with the abnormal vehicle to the lowest to ensure the normal traffic of vehicles in other lanes. Here, a vehicle will be labeled as an abnormal vehicle if the passing time $t_e$ of the vehicle through the conflict point is larger than:
\begin{equation}
t_e > \frac{A_l}{\frac{v_{limit}}{2}}+ \sigma,
\end{equation}
where $\sigma$ is the time constant of abnormal judgment.

Based on real-world human driving habits, at unsignalized intersections, HV behaviors can be categorized into two primary types: following the preceding vehicle and free-driving in the absence of preceding vehicles \cite{YU2021103101}. Based on the previous study \cite{YU2021103101}, we developed rules for CAVs, CHVs, and HVs to pass through unsignalized intersections safely:
\begin{enumerate}[1)]
\item When there is a trajectory conflict between the CAVs or CHVs and the HVs while passing through an intersection, the CAVs and CHVs will give priority to the HVs to ensure safe passage through the intersection. The CAV is allowed to enter the intersection conflict zone with one HV at the same time by enabling the conflict resolution mode.
\item When two free HVs from conflicting directions reach the intersection simultaneously, they adhere to the right-hand rule to pass through the intersection. 
\end{enumerate}

The HV behavior model is modeled by referring to the previous study \cite{kanagaraj2013evaluation}.
When there is a vehicle in front of the HV, the HV has to maintain a safe distance to avoid collision with the preceding vehicle, and thus, the HV needs to be maintained under a safe speed in real time:
\begin{equation}
v_{safe}(t) = v_f(t) + \frac{g(t)-v_f(t)\tau}{\frac{v}{a_d}+\tau},
\end{equation}
where $v_f(t)$ is the speed of the preceding vehicle at time $t$, $g(t)$ is the distance between the two vehicles at time $t$, $\tau$ is the reaction time of the human driver, and $a_d$ is the maximum deceleration of the vehicle. In reality, the target speed of the vehicle should not only consider the above situation, but also consider the dynamic characteristics of the vehicle itself:
\begin{equation}
v_{des}(t) = min(v_{safe}(t), v(t-1)+a, v_{max}),
\end{equation}
where $v_{max}$ denotes the maximum speed limit when the vehicle passes through the intersection.

To model the uncertainty of human driving behavior more accurately, a random error $\epsilon$ is introduced here to portray the relative instability of human drivers' maneuvers:
\begin{equation}
v(t) = max(0, rand(v_{des}(t)-\epsilon a, v_{des}(t)+\epsilon a)).
\end{equation}

In the case of free-driving in the absence of preceding vehicles, the speed of the vehicle depends primarily on the speed limit $v_{limit}$, the speed of the yielding vehicle $v_{yielding}$, and the driver's target speed $v_{target}$.

\subsection{Right of Way Management Model}
The most significant difference between an intersection with and without signal lamps is that the former allocates the right of way for each lane, while the latter allocates the right of way for each vehicle individually. Here, the factors that affect the right of way of each vehicle include vehicle priority, trajectory conflicts, and vehicle type. To save computing resources, the ICU divides all vehicles $S$ within the intersection into three groups. The first group $S_1$ is the set of vehicles that have passed through the conflict area, which does not need intersection management; the second group $S_2$ is the set of vehicles that have not passed through the conflict area but have obtained the right of way; the third group $S_3$ is the set of vehicles that have not obtained the right of way. The right of way management only needs to determine which vehicles in $S_3$ can get the right of way.

\begin{figure}[!t]
\centering
\includegraphics[width=3.3in]{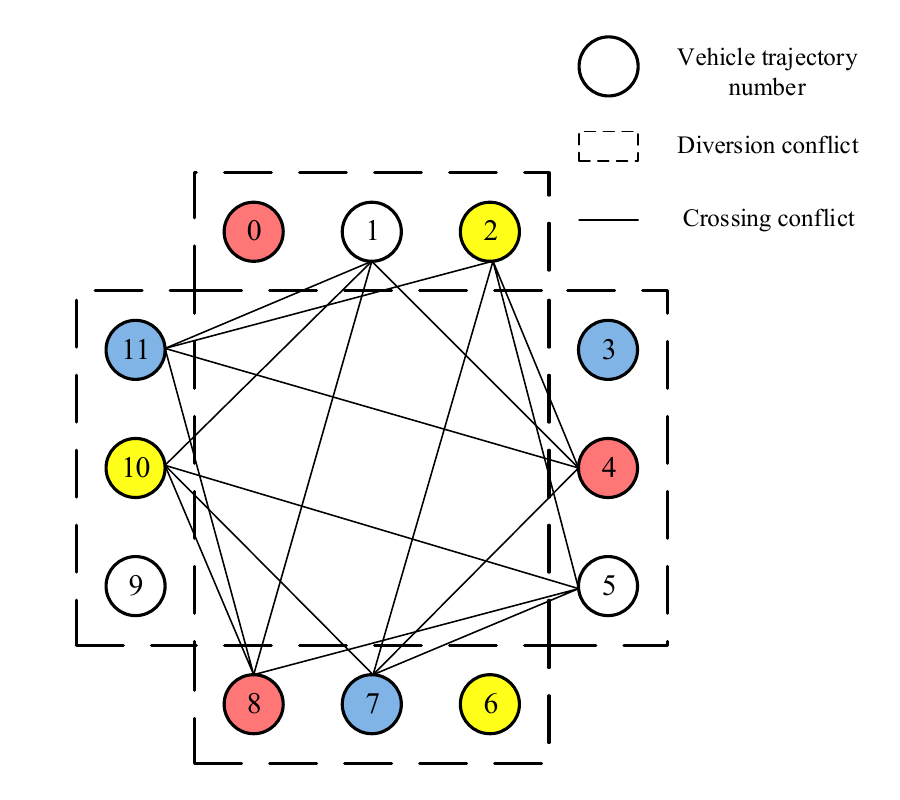}
\caption{Vehicle trajectory conflict relationship diagram at intersections.}
\label{fig_Vehicle_trajectory_conflict}
\end{figure}

In the same lane, the vehicle in the front must have priority over the vehicle in the back. Therefore, the third group $S_3$ can be divided into four priority queues ($Q_1, Q_2, Q_3, Q_4$) according to the four entry lanes, where $Q_i$ corresponds to the $i$-th lane and ranks the priority of vehicles in the $i$-th lane from small to large according to the time stamp of the vehicle entering the intersection. Every time when the right of way is allocated, the ICU only needs to consider the vehicles at the top of the four queues $Q_1[0], Q_2[0], Q_3[0], Q_4[0]$, which are classified as set $Q_p$. For the four vehicles at the top of the queue, sort them according to their priorities according to the FCFS policy and check whether the vehicles meet the following conditions for obtaining the right of way. The objective is to find one vehicle that meets the conditions to obtain the right of way. According to the differences between CAVs, CHVs, and HVs, the conditions are set respectively:
\subsubsection{Conditions for CHVs to obtain the right of way} There is no trajectory conflict with the vehicles in the second set $S_2$, and there is no trajectory conflict with the vehicles with higher priority in the third set $S_3$.
\subsubsection{Conditions for CAVs to obtain the right of way} There can be a trajectory conflict with at most one vehicle in the second set $S_2$, and there is no trajectory conflict with the vehicles with higher priority in the third set $S_3$.
\subsubsection{Conditions for HVs to obtain the right of way} When there are preceding vehicles, HVs follow them through the intersection. When there are no preceding vehicles, HVs adhere to the right-hand rule to pass through the intersection.

For trajectory conflict in the typical unsignalized intersection as shown in Fig. 2, straight-through vehicles in entry area $A_3$ have crossing conflicts with straight-through vehicles and left-turn vehicles in entry area $A_2$, and have merging conflicts with right-turn vehicles in entry area $A_2$. Here, pairwise conflicting relationships are represented by an undirected conflict graph $\mathcal{G} = (\mathcal{V}, \mathcal{E})$, which will be used to sort vehicles without conflicts that can pass the intersection at the same time. Fig. 4 shows the vehicle trajectory conflict relationship diagram at the intersection. The number in the circle in Fig. 4 represents the trajectory number, where vehicle trajectory numbers 0, 3, 6, 9 represent right-turn movements; vehicle trajectory numbers 1, 4, 7, 10 represent through movements; vehicle trajectory numbers 2, 5, 8, 11 represent left-turn movements. And the circle with the same color represents the existence of merging conflicts. The trajectories in Fig. 4 can be regarded as a set of nodes $\mathcal{V} = \{v_0, v_1, v_2, \cdots, v_{n-1}\}$. And the crossing conflict between two trajectories can be represented by an edge, as indicated in Fig. 4. Here, $n=12$ for the four-way intersection, and the set of the edges is $\mathcal{E} = \{e_0, e_1, e_2, \cdots, e_{m-1}\}$, with $m=16$. 


When a vehicle enters the intersection, it should send its own position, speed, acceleration, and other kinematics information to the ICU, including the type of vehicle $C(V)$ ($C(V) == 1$ means CHV, and $C(V) == 0$ means CAV), the time stamp of entering the intersection and the expected driving direction (going straight, turning right and turning left). After the ICU collects all the vehicles' information, it will execute the HPQ algorithm once in each control cycle to grant at most one vehicle the right of way. 
Once a vehicle obtains the right of way, the algorithm will update the status of all sets of vehicles and priority queues. Here, a control cycle means that the HPQ algorithm performs one cycle calculation on all vehicles in the priority queue $Q_p$ and outputs the result. The length of the control cycle will be changed according to the traffic flow at the intersection. 

For the HPQ algorithm shown in Algorithm 1, the parameters $N_1$ and $N_2$ need to be initialized, where $N_1$ represents the number of trajectory conflicts between the vehicle $V$ in the priority queue $Q_p$ and other vehicles in the set $S_2$ that have been granted the right of way, and $N_2$ represents the number of trajectory conflicts between the vehicle $V$ in the priority queue $Q_p$ and the other vehicles in the set $S_3$ that have not obtained the right of way. Vehicle $V$ is the vehicle to be checked by the HPQ algorithm to determine whether it can be granted the right of way. The priority value $P(V)$ can be calculated by the priority management model.   

The HPQ algorithm will find the vehicle that can be granted the right of way in the priority queue $Q_p$. First, for each candidate vehicle $V$ in $Q_p$, check the vehicle trajectory conflict and calculate the values of $N_1$ and $N_2$ according to the vehicle trajectory conflict relationship diagram in Fig. 4. If there is a trajectory conflict between the vehicle $V$ in the priority queue $Q_p$ and the other vehicles in $S_2$, the ID of the conflict vehicle will be recorded in the trajectory conflict variable $I_c$ and the value of $N_1$ will increase by 1. Then find the vehicle $\widetilde{V}$ in the set $S_3$ whose priority value $P(\widetilde{V})$ is smaller than the priority value $P(V)$ of vehicle $V$ in the priority queue $Q_p$, where $S_3$ is the set of vehicles that have not obtained the right of way. If there is a trajectory conflict between the vehicle $V$ in the priority queue $Q_p$ and the other vehicle $\widetilde{V}$, the value of $N_2$ will increase by 1. Finally, judge whether the vehicle $V$ in the priority queue $Q_p$ is a CHV or CAV. If the vehicle $V$ in the priority queue $Q_p$ is a CHV and satisfies the conditions for CHVs to obtain the right of way, that is, there is no trajectory conflict between the vehicle $V$ in the priority queue $Q_p$ and the other vehicles in the sets $S_2$ and $S_3$, the vehicle $V$ in the priority queue $Q_p$ is granted the right of way. If the vehicle $V$ in the priority queue $Q_p$ is a CAV and satisfies the conditions for CAVs to obtain the right of way, that is, the number of trajectories conflict between the vehicle $V$ in the priority queue $Q_p$ and the other vehicles in the set $S_2$ do not exceed one and there is no trajectory conflict between the vehicle $V$ in the priority queue $Q_p$ and the other vehicles in the set $S_3$, vehicle $V$ is granted the right of way. If the above conditions are not met, no vehicle is granted the right of way, and the next control cycle is performed. 

When a vehicle is granted the right of way, the ICU will issue an instruction to allow the vehicle to pass.
Finally, further conflict checks are carried out according to the real-time location of the vehicles at the intersection. If there is a trajectory conflict that requires to be dealt with, the ICU will also send the information of the trajectory conflict to the conflict vehicles. In addition, CAVs can obtain information about other vehicles through V2V communication so as to improve the accuracy of low-level vehicle planning and control.

\begin{figure}[!t]
	\label{alg1}
	\renewcommand{\algorithmicrequire}{\textbf{Input:}}
	\renewcommand{\algorithmicensure}{\textbf{Output:}}
	\removelatexerror
	\begin{algorithm}[H]
		\caption{HPQ Algorithm}
		\begin{algorithmic}[1]
		
		\REQUIRE $U$
		\ENSURE $I_p$, $I_c$
        \STATE Initialize
        \STATE $N_1$ = 0, $N_2$ = 0, [$I_p$, $I_c$] = [` ', ` ']
        \FOR{$V$ in $Q_p$}
        \FOR{$\widetilde{V}$ in $S_2$}
        \IF{$T_r(V)$ $\cap$ $T_r(\widetilde{V})$ ! = $\emptyset$}
        \STATE $N_1$ + = 1
        \STATE $I_c$ = $\widetilde{V}$
        \ENDIF
        \ENDFOR
        
        \FOR{$\widetilde{V}$ in $S_3$}
        \IF{$P(\widetilde{V})$ $\geq$ $P(V)$}
        \STATE continue
        \ENDIF
        \IF{$T_r(V)$ $\cap$ $T_r(\widetilde{V})$ ! = $\emptyset$}
        \STATE $N_2$ + = 1
        \ENDIF
        \ENDFOR
        
        \IF{$C(V)$ == 1}
        \IF{$N_1$ + $N_2$ == 0}
        \RETURN [$V$, ` ']
        \ENDIF
        \ELSE
        \IF{$N_1 \leq 1$ \AND $N_2 = 0$}
        \RETURN [$V$, $I_c$]
        \ENDIF
        \ENDIF
        \ENDFOR
        \RETURN [` ', ` ']
		\end{algorithmic}
	\end{algorithm}
\end{figure}

The time complexity of the right of way management algorithm is $O(n)$, where $n$ is the total number of vehicles at the intersection. In HPQ Algorithm 1, a vehicle information struct ($U$), five priority queues ($Q_1$, $Q_2$, $Q_3$, $Q_4$, $Q_p$), and four vehicle sets ($S$, $S_1$, $S_2$, $S_3$) need to be updated in each control cycle. The priority queue is implemented by the heap. The time complexity of building a heap is $O(n)$, and the time complexity of heap insertion and deletion is $O(\log n)$. After that, to check whether a vehicle conflicts with another vehicle, it needs to query the adjacency matrix that represents the conflict relationship, and the time complexity of this operation is $O(1)$. Then, it needs to query all the corresponding sets $S_2$ and $S_3$, and the time complexity is $O(n)$. Because there are only four vehicles in $Q_p$ at most, the time complexity of querying all vehicles in $Q_p$ is $O(n)$. Therefore, the overall time complexity is $O(n)$, which guarantees the efficiency of real-time control for the right of way can be achieved. 

According to the designed HPQ algorithm, vehicles with low priority may get the right of way earlier than vehicles with high priority. To take advantage of the intersection conflict area as efficiently as possible, vehicles with low priority can obtain the right of way as long as they do not affect the vehicles with high priority and vehicles that have already obtained the right of way. This means that vehicles from different entry areas of the intersection can pass the intersection simultaneously. For example, the vehicle with trajectory number 11 in Fig. 4 arrives at the intersection first and gets the highest priority, and then vehicles with trajectory numbers 0 and 5 also enter the intersection. Although the vehicle with trajectory number 11 has the highest priority, vehicles with trajectory numbers 0 and 5 can also obtain the right of way, and these three vehicles can pass the intersection simultaneously. 
Moreover, the CAVs are allowed to obtain the right of way in the case of conflicting with one vehicle that has been granted the right of way, since CAVs can solve the vehicle conflict through the designed low-level vehicle planning and control algorithm to ensure the safety of vehicles.

\begin{figure}[!t]
\centering
\includegraphics[width=3.3in]{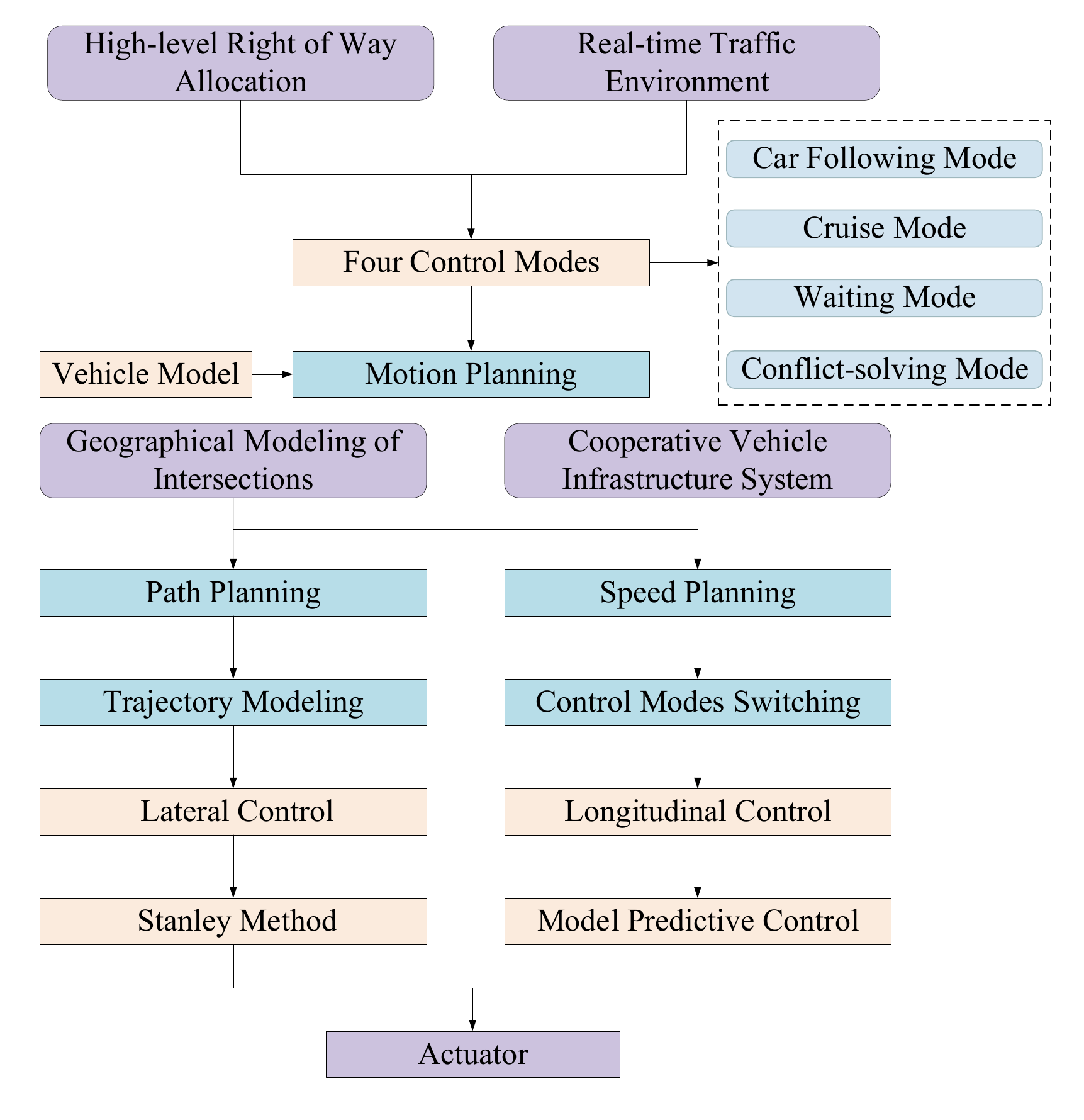}
\caption{The whole procedure of vehicle planning and control. Based on the right of way allocation results calculated by the HPQ algorithm and the real-time traffic environment, the control modes of vehicles are calculated by the control modes switching algorithm. Then, motion planning is carried out, in which path planning and speed planning run in parallel to obtain the best lateral and longitudinal control scheme. All these control actions are finally output to the actuator for execution.}
\label{fig_vehicle_planning_and_control}
\end{figure}

\section{VEHICLE PLANNING AND CONTROL ALGORITHM DESIGN}
In the previous section, we developed the HPQ algorithm, which combines the differences between CAVs, CHVs, and HVs to arrange the corresponding right of way. In this section, we focus on vehicle planning and control for CAVs since the ICU has high computing power and can obtain the global information of the traffic participants near the intersection. 

According to the driving situation of CAVs at the intersection, we introduce four possible decision making modes for CAVs: the first mode is to follow the preceding vehicle; the second mode is autonomous cruise; the third mode is to stop before the stop line to wait for the right of way; and the last mode is to resolve the trajectory conflicts. In the following text, the corresponding planning and control methods for the four decision making modes of CAVs will be designed in detail. The motion planning problem is divided into two subproblems, including path planning and speed planning, where path planning is constrained by the layout of the road and the geographical model of the intersection, and speed planning enables passing through the intersection quickly, safely, and economically. By using model predictive control, CAVs can switch undisturbedly to the corresponding decision making mode according to the current traffic environment. The whole procedure of vehicle planning and control is illustrated in Fig. 5. Based on the right of way allocation results calculated by the HPQ algorithm and the real-time traffic environment, the control modes of vehicles are calculated by the control modes switching algorithm. Then, motion planning is carried out, in which path planning and speed planning run in parallel to obtain the best lateral and longitudinal control scheme. All these control actions are finally output to the actuator for execution.    
\subsection{Path Planning Model}
For a typical four-way unsignalized intersection, as shown in Fig. 6, the intersection trajectory is modeled, and the intersection coordinate system, lane coordinate system, and vehicle coordinate system are established to carry out vehicle path planning. The intersection coordinate system $\xi$ originates from the center point $O$ of the intersection, the Y-axis points north, and the X-axis points east. The lane coordinate system $\xi_i, i \in \{1, 2, 3, 4\}$ takes the midpoint of the starting position of each lane in the entry area as the origin, where A-axis is along the center line of the lane, and B-axis is obtained by rotating A-axis 90 degrees counterclockwise. The vehicle coordinate system $\xi_v$ takes the vehicle centroid as the origin, where M-axis is the direction of vehicle heading, and N-axis is obtained by rotating M-axis 90 degrees counterclockwise.

\begin{figure}[!t]
\centering
\includegraphics[width=3.3in]{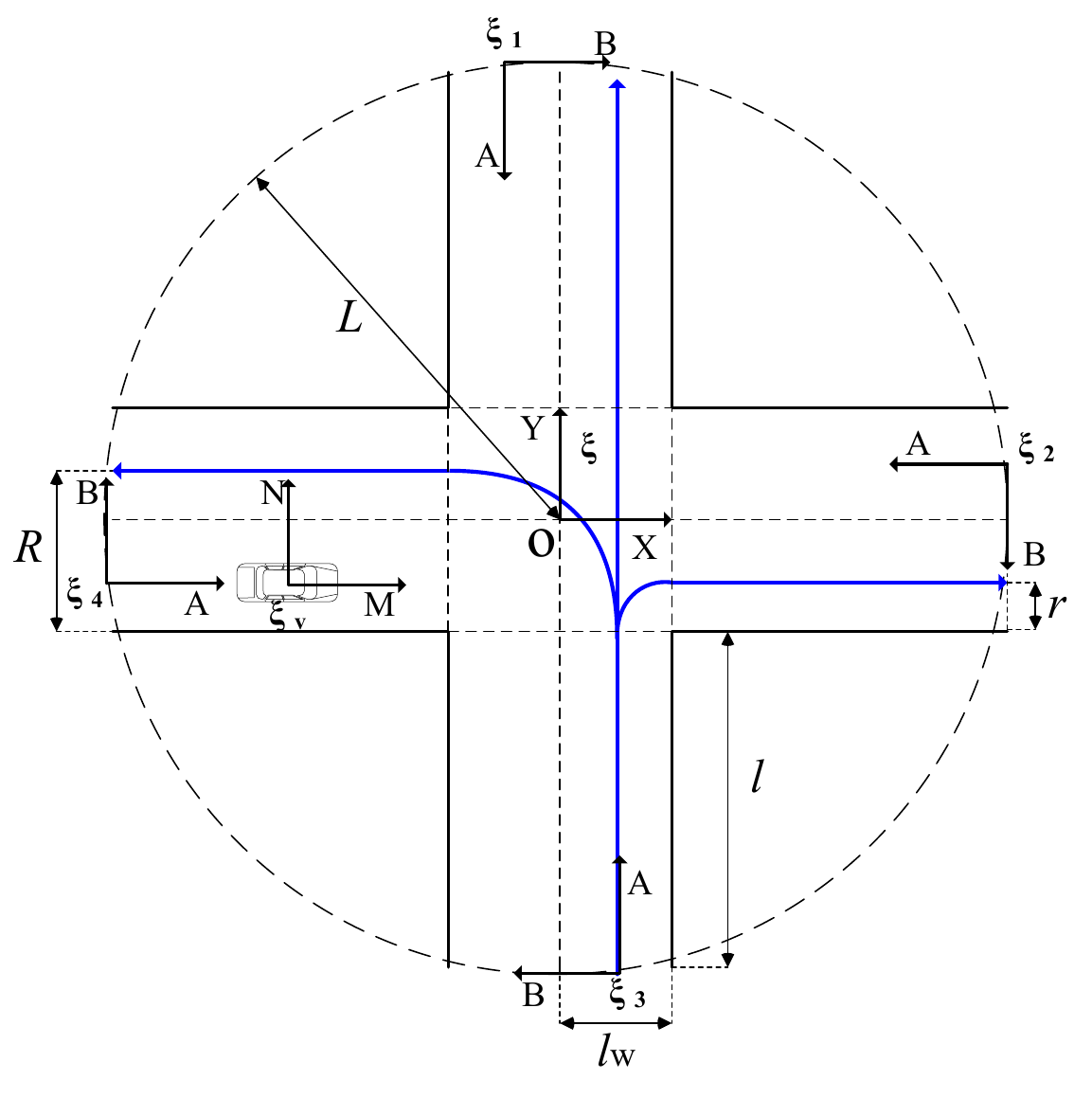}
\caption{Intersection coordinate system and trajectories of vehicles.}
\label{fig_Vehicle_coordinate_system}
\end{figure}





The trajectory tracking of CAVs can be solved by vehicle lateral control \cite{8932661, 7378957, 8868658}. For convenience, the Cartesian coordinate system is transformed into the coordinate system for vehicle path tracking, and the corresponding descriptions for left turn trajectory, straight trajectory, and right turn trajectory are formulated as follows. The coordinate of the CAVs in the lane coordinate system is denoted as $(\alpha_p, \beta_p)$. The distance traveled by the CAVs is denoted as $s$. The heading angle of the CAVs is denoted as $\theta$. The left turn trajectory is shown in Equations (7) and (8). The straight trajectory is shown in Equations (9) and (10). The right turn trajectory is shown in Equations (11) and (12).
\begin{equation}  {s = }
\left\{
  \begin{array}{ll}
\alpha_p, &\beta_p = 0, 0 \le \alpha_p \le l,\\
l + R \times \arctan \frac{\alpha_p-l}{R-\beta_p}, & l < \alpha_p < l+R, 0 < \beta_p < R,\\
l + \frac{1}{2}\pi R + \beta_p -R, & \alpha_p = l+R, R \le \beta_p \le R+l.
  \end{array}
\right.
\end{equation}
\begin{equation}  {\theta = }
\left\{
  \begin{array}{ll}
0, &\beta_p = 0, 0 \le \alpha_p \le l,\\
\arctan \frac{\alpha_p-l}{R-\beta_p}, & l < \alpha_p < l+R, 0 < \beta_p < R,\\
90, & \alpha_p = l+R, R \le \beta_p \le R+l.
  \end{array}
\right.
\end{equation}
\begin{equation}  
s = \alpha_p,  \ \  0 \le \alpha_p \le 2L, \beta_p = 0.
\end{equation}
\begin{equation}  
\theta = 0,  \ \ \  0 \le \alpha_p \le 2L, \beta_p = 0.
\end{equation}
\begin{equation}  {s = }
\left\{
  \begin{array}{ll}
\alpha_p, &\beta_p = 0, 0 \le \alpha_p \le l,\\
l + r \times \arctan \frac{\alpha_p-l}{r+\beta_p}, & l < \alpha_p < l+r, -r < \beta_p < 0,\\
l + \frac{1}{2}\pi r - \beta_p -r, & \alpha_p = l+r, -l-r \le \beta_p \le -r.
  \end{array}
\right.
\end{equation}
\begin{equation}  {\theta = }
\left\{
  \begin{array}{ll}
0, &\beta_p = 0, 0 \le \alpha_p \le l,\\
-\arctan \frac{\alpha_p-l}{r+\beta_p}, & l < \alpha_p < l+r, -r < \beta_p < 0,\\
-90, & \alpha_p = l+r, -l-r \le \beta_p \le -r.
  \end{array}
\right.
\end{equation}
where $l$ is the length of the entry area; $l_w$ is the width of the lane; $L$ is the radius of the control range of the intersection area; $R$ and $r$ are the radius of turning left and turning right respectively, and meet the constrain $R + r = 2 l_w$.

\subsection{Speed Planning Model}
Since CAVs need to make corresponding decisions according to different environments, four control modes are designed to ensure the safety of vehicles and improve traffic efficiency. Each control mode includes a corresponding speed plan. The control modes of CAVs are listed as follows:
\begin{itemize}
\item Car following mode: Keep a safe distance from the preceding vehicle and follow it.
\item Cruise mode: Keep the vehicle at a constant speed.
\item Waiting mode: Slow down until the vehicle is stopped before the stop line.
\item Conflict-solving mode: Resolve vehicle trajectory conflicts when passing through conflict areas.
\end{itemize}

To improve the conflict resolution ability of CAVs and ensure the safety of vehicles, the real-time switching of four vehicle control modes needs to be guaranteed. Define a bool type variable $B$ to denote whether the ego vehicle $V$ obtains the right of way at the intersection. $I_c$ represents the vehicle's ID with trajectory conflicts with the ego vehicle $V$, and $I_f$ represents the ID of the preceding vehicle of ego vehicle $V$. Define $S_V(E)$ and $S_V(stopline)$ to denote the position of vehicle $E$ and the position of the stop line in the vehicle coordinate system corresponding to the ego vehicle $V$, respectively. And $\epsilon$ is the distance constant. 

We develop an undisturbed switching algorithm for the control modes by combining the real-time traffic environment around the ego vehicle $V$. 
First, judge whether each vehicle has obtained the right of way according to the high-level HPQ algorithm. For vehicles without obtaining the right of way: if there is a preceding vehicle, the control mode is switched to the car following mode; if there is no vehicle ahead and vehicle $V$ reaches near the stop line, the control mode is switched to the waiting mode; otherwise, the control mode is switched to cruise mode. 

For vehicles with the right of way: if it can form a vehicle formation with the preceding vehicles to travel together without interference from other vehicles, the control mode is switched to the car following mode; if the vehicle formation may be interfered by other vehicles, the control mode is switched to conflict-solving mode; otherwise, the control mode is switched to cruise mode. Define $ind(min(d_1, d_2))$ as an index when the minimum function gets the minimum value. The value of $ind(min(d_1, d_2))$ is 0 when $d_1 \le d_2$, otherwise it is 1. The principles of switching between the four control modes are shown in Algorithm 2.

\begin{figure}[!t]
	\label{alg2}
	\renewcommand{\algorithmicrequire}{\textbf{Input:}}
	\renewcommand{\algorithmicensure}{\textbf{Output:}}
	\removelatexerror
	\begin{algorithm}[H]
		\caption{Switching Algorithm for Control Modes}
		\begin{algorithmic}[1]
		
		\REQUIRE $B$, $I_c$, $I_f$
		\ENSURE $M(V)$
		\FOR{$V$ in $S$}
        \IF{$B$ = = $false$}
        \IF{$I_f$ ! = ` '}
        \STATE $M(V)$ = `Car following'
        \ELSIF{$S_V(stopline)$ - $S_V(V)$ \textless     $\epsilon$}
        \STATE $M(V)$ = `Waiting'
        \ELSE
        \STATE $M(V)$ = `Cruise'
        \ENDIF

        \ELSE
        \IF{$I_f$ ! = ` ' \AND ($I_c$ = = ` ' \OR $I_c$ ! = ` ' \AND ind(min($S_V(I_c)$ - $S_V(V)$, $S_V(I_f)$ - $S_V(V)$))) = = 1}
        \STATE $M(V)$ = `Car following'
        \ELSIF{$I_c$ ! = ` ' \AND ($I_f$ = = ` ' \OR $I_f$ ! = ` ' \AND ind(min($S_V(I_c)$ - $S_V(V)$, $S_V(I_f)$ - $S_V(V)$))) = = 0}
        \STATE $M(V)$ = `Conflict-solving'
        \ELSE
        \STATE $M(V)$ = `Cruise'
        \ENDIF
        \ENDIF
        \ENDFOR
		\end{algorithmic}
	\end{algorithm}
\end{figure}

Let $s$ and $v$ be the current position and velocity of vehicle $V$ in the corresponding lane coordinate system. For the four control modes of CAVs, different reference positions $s_{ref}$ and reference speeds $v_{ref}$ are designed respectively, as shown in Equations (13-16).

\begin{figure}[!t]
\centering
\includegraphics[width=3.5in]{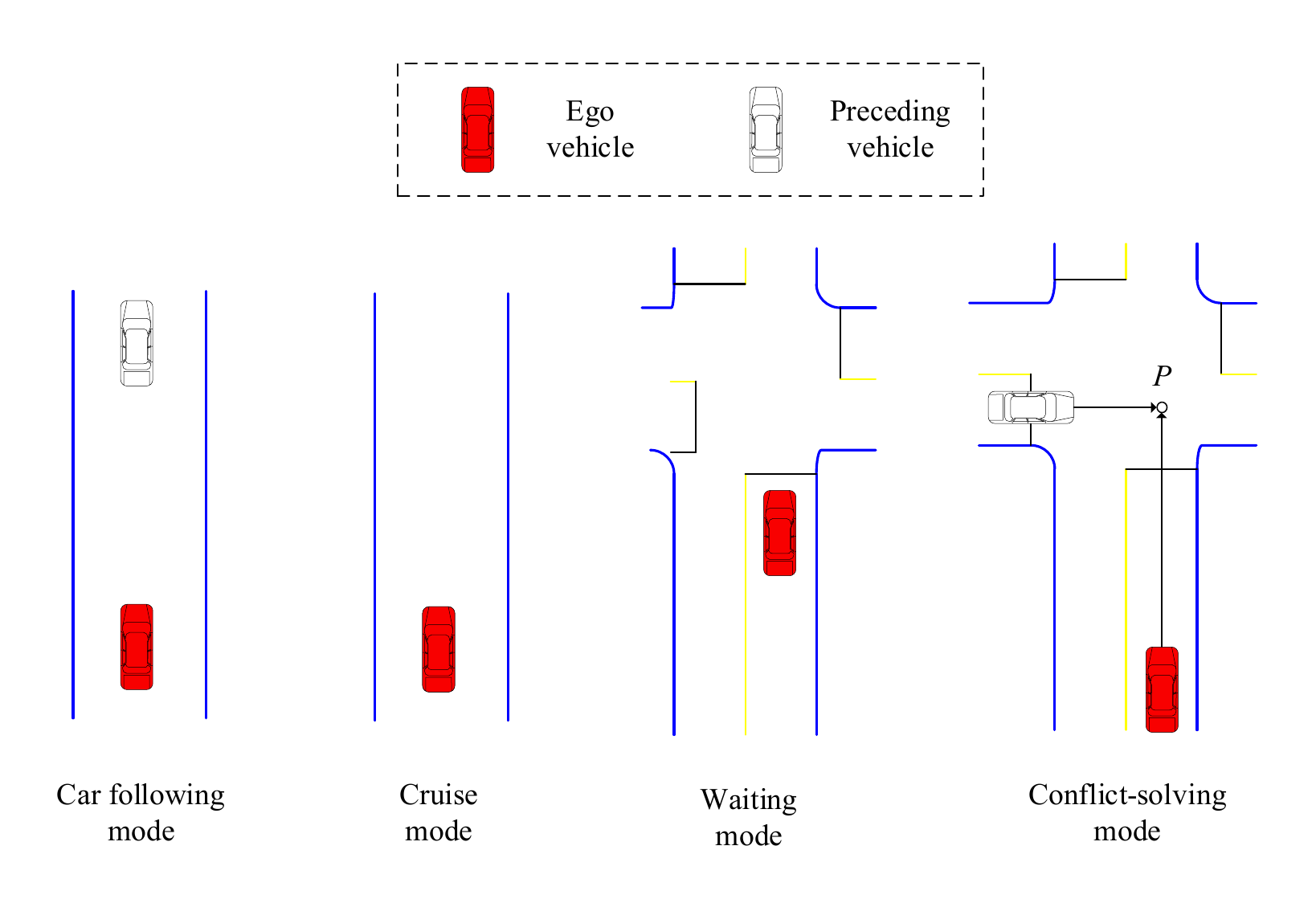}
\caption{Four control modes of CAVs.}
\label{fig_Vehicle_control_mode}
\end{figure}

\begin{itemize}
\item Car following mode: 
\end{itemize}
\begin{equation}  
\left\{
  \begin{array}{ll}
s_{ref} = s_f - d_{min} - l_V - hv, \\
v_{ref} = v_f,\\
  \end{array}
\right.
\end{equation}
where $s_f$ and $v_f$ are the position and speed of the preceding vehicle, respectively; $d_{min}$ is the minimum safety distance between the ego vehicle and the preceding vehicle; $h$ is the minimum travel time from the rear of the preceding vehicle to the front of the following vehicle; and $l_V$ is the length of vehicle $V$.

\begin{itemize}
\item Cruise mode:
\end{itemize}
\begin{equation}  
\left\{
  \begin{array}{ll}
s_{ref} = s, \\
v_{ref} = v_{tar},\\
  \end{array}
\right.
\end{equation}
where $v_{tar}$ is target speed of vehicle $V$.

\begin{itemize}
\item Waiting mode:
\end{itemize}
\begin{equation}  
\left\{
  \begin{array}{ll}
s_{ref} = s_{stopline} - l_{tar}, \\
v_{ref} = 0,\\
  \end{array}
\right.
\end{equation}
where $s_{stopline}$ is the position of stop line and $l_{tar}$ is the target distance from vehicle centroid to the stop line.

\begin{itemize}
\item Conflict-solving mode:
\end{itemize}
\begin{equation}  
\left\{
  \begin{array}{ll}
s_{ref} = s_{p} - \widetilde{s_p} +\widetilde{s_f} -d_{min} -l_V -hv, \\
v_{ref} = v_f,\\
  \end{array}
\right.
\end{equation}
where $s_{p}$ is the position of the conflict point $P$ in the lane coordinate system corresponding to vehicle $V$; $\widetilde{s_p}$ is the position of the conflict point $P$ in the lane coordinate system corresponding to the vehicle that conflicts with vehicle $V$, and $\widetilde{s_f}$ is the position of the vehicle $V$ in the lane coordinate system corresponding to the vehicle that conflicts with vehicle $V$.

The conflict-solving mode uses the idea of virtual formation \cite{medina2017cooperative} to solve the problem of vehicle conflict, as shown in Fig. 8. To make the preceding vehicle and the ego vehicle pass through the conflict point $P$ safely, we improve the speed planning model by introducing four control modes of undisturbed switching and virtual formation. Assuming that the distance between the ego vehicle and conflict point $P$ is $\Delta s_b$ and the distance between the preceding vehicle and conflict point $P$ is $\Delta s_f$. In order to make the time interval between the ego vehicle and the preceding vehicle reaching conflict point $P$ equal to $H$, which is greater than the minimum safety time headway, the virtual formation is used for position mapping to calculate the relative position between conflicting vehicles. The position of the preceding vehicle can be mapped from the original lane to the lane the ego vehicle is in, and the relative distance between the preceding vehicle and the conflict point $P$ remains unchanged, which meets $\Delta s_f' = \Delta s_f$, as illustrated in Fig. 8. As long as the ego vehicle and the virtual preceding vehicle maintain time headway $H$, both the ego vehicle and the preceding vehicle can pass through the intersection safely and quickly. Different types of vehicles have different time headway $H$ to ensure the vehicle's safety to the greatest extent \cite{8884686}.


\begin{figure}[!t]
\centering
\includegraphics[width=2.8in]{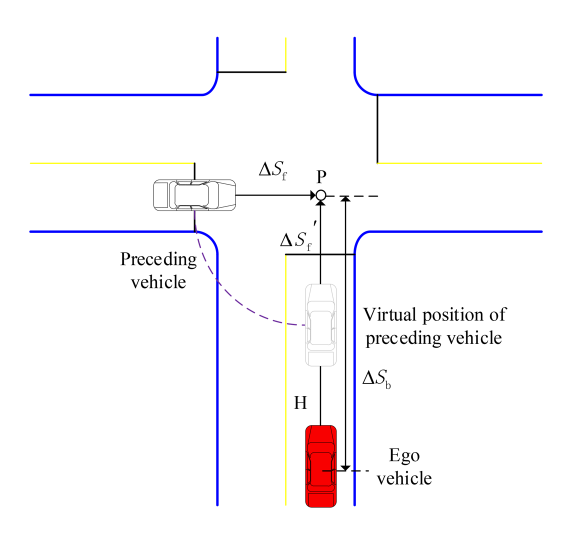}
\caption{Schematic diagram of virtual formation method for collision avoidance.}
\label{fig_Vehicle_virtual_formation}
\end{figure}

\subsection{Vehicle Control based on Model Predictive Control}
\begin{itemize}
\item Lateral control: 
\end{itemize}

The vehicle steering model uses the Stanley method \cite{4282788}, as shown in Fig. 9. To obtain the steering model, it is essential to determine the closest point from the center point of the front axle to the target trajectory. The blue curve in Fig. 9 is the target trajectory obtained by the path planning model; $(\alpha_c, \beta_c)$ is the closest point; $e_{sc}$ is the offset between the center point of the front axle and the closest point $(\alpha_c, \beta_c)$; and $\theta_e$ is the angle between the vehicle heading and the tangent to the trajectory of the closest point. The steering angle $\delta$ of the front wheel can be calculated as Equation (17).
\begin{equation}  
\delta = \theta_V -\theta_P + \arctan (\frac{k_f e_{sc}}{v}),
\end{equation}
where $\theta_V$ is the heading angle of vehicle $V$; $\theta_P$ is the tangent angle at point $(\alpha_c, \beta_c)$ relative to the target trajectory; $k_f$ is the proportional parameter; and $v$ is the speed of vehicle $V$.

\begin{itemize}
\item Longitudinal control:
\end{itemize}

For longitudinal control, the speed planning model is controlled by the model predictive controller designed in \cite{he2020robust}. Define the longitudinal position error of vehicle $V$ as $e_{sf} = s_{ref} - s$ and the speed error as $e_v = v_{ref} - v$. Error state equations of CAVs in continuous time domain are shown in Equations (18-19), which is utilized in \cite{he2020robust} to establish the state space model of longitudinal dynamics of vehicles.
\begin{equation}  
\dot{x}(t) = \widetilde{\Xi} x(t) + \widetilde{\Psi}_1 \tau(t) +\widetilde{\Psi}_2 \omega(t), \\
\end{equation}
\begin{equation}
\widetilde{\Xi} = 
\left[                
  \begin{array}{cc}   
    0 & 1\\  
    0 & 0\\  
  \end{array}
\right],           \ \       
\widetilde{{\Psi}}_1 = 
\left[                
  \begin{array}{c}   
    - \widetilde{h}\\  
    - 1\\  
  \end{array}
\right],          \ \        
\widetilde{{\Psi}}_2 = 
\left[                
  \begin{array}{c}   
      0\\  
      1\\  
  \end{array}
\right],          \ \        
\end{equation}
where $x = [e_{sf}, e_v]^T$, $\tau = a$, $\omega = a_t$. $a$ is the acceleration of vehicle $V$, and $a_t$ is the acceleration of other vehicles, including the preceding vehicle or the vehicle that conflicts with vehicle $V$. $a_t = 0$ means no preceding vehicle or vehicle conflicts with vehicle $V$. If the control mode of vehicle $V$ is car following mode or conflict-solving mode, $\widetilde{h} = h$. Otherwise, $\widetilde{h} = 0$. 

\begin{figure}[!t]
\centering
\includegraphics[width=2.8in]{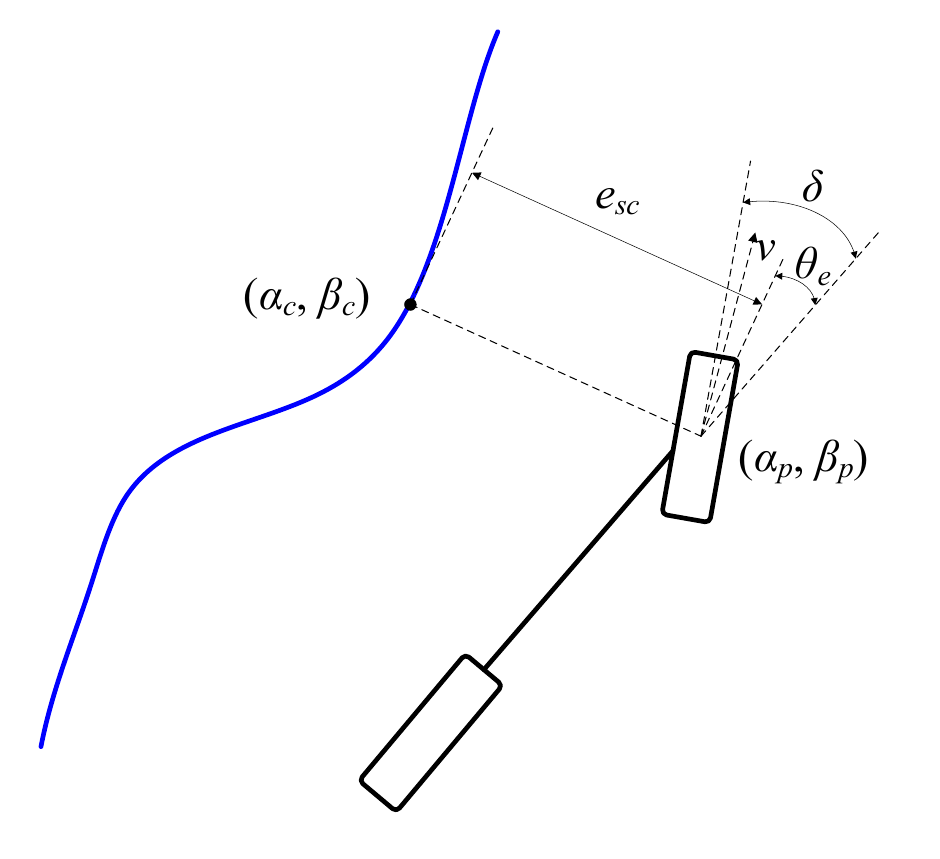}
\caption{The Stanley method used for lateral control.}
\label{fig_stanley_model}
\end{figure}

Since the update of the vehicle state quantity $x$ is discrete, the Equations (18-19) are transformed into discrete linear state equations as shown in Equations (20-21).
\begin{equation}  
{x}(\kappa+1) = \Xi x(\kappa) + {\Psi}_1 \tau(\kappa) + {\Psi}_2 \omega(\kappa), \\
\end{equation}

\begin{equation}
{\Xi} = 
\left[                
  \begin{array}{cc}   
    1 & \varphi\\  
    0 & 1\\  
  \end{array}
\right],           \ \       
{{\Psi}}_1 = 
\left[                
  \begin{array}{c}   
    - \widetilde{h} \varphi -\frac{1}{2} \varphi^2\\  
    - \varphi\\  
  \end{array}
\right],          \ \        
\Psi_2 = 
\left[                
  \begin{array}{c}   
      \frac{1}{2} \varphi^2\\  
      \varphi\\  
  \end{array}
\right],          \ \        
\end{equation}
where $\varphi$ is the control cycle. We use the idea of rolling optimization in model predictive control to realize the longitudinal control of the CAVs. The overall objective is to minimize the total deviation between the vehicle speed $v$ and the reference speed $v_{ref}$ in a finite period of time in the future. To make the behaviors of the CAVs smoother and to improve the driving comfort of the CAVs, we further introduce acceleration constraints \cite{chen2020hierarchical}. The objective function and constraints of the proposed algorithm are as follows:
\begin{equation}
\begin{aligned} \label{P}
&\min_\tau \sum_{\eta=1}^{N_p} \left \| x(\kappa + \eta | \kappa)  \right \|_{\Theta}^2 + \sum_{\eta=0}^{N_c-1} \left \| \tau(\kappa + \eta | \kappa)  \right \|_{\Phi}^2, \\
s.t. \;  &x_{min} \le x(\kappa + \eta | \kappa) \le x_{max},  \\
     &\tau_{min} \le \tau(\kappa + \eta | \kappa) \le \tau_{max},  \\  
     &\Delta \tau_{min} \le \tau(\kappa + \eta | \kappa) - \tau (\kappa + \eta - 1 | \kappa)\le \Delta \tau_{max},  \\
     &a_{d} \le a(\kappa + \eta | \kappa) \le a_{com},  
\end{aligned}
\end{equation}
where $N_p$ is the predictive step length; $N_c$ is the control step length; $\Theta$ is the state weight matrix; $\Phi$ is the input weight matrix; $x_{min}$ and $x_{max}$ are the lower limit of the state quantity and the upper limit of the state quantity respectively; $\tau_{min}$ and $\tau_{max}$ are the lower limit of the control quantity and the upper limit of the control quantity respectively; $\Delta \tau_{min}$ and $\Delta \tau_{max}$ are the lower limit of the increment of control quantity and the upper limit of the increment of control quantity respectively; $a_d$ is the maximum deceleration without loss of comfort, which is determined to be $-4m/s^2$ by drawing on \cite{chen2020hierarchical}. $a_{com}$ is the maximum acceleration without loss of comfort, which is determined to be $2m/s^2$ by drawing on \cite{chen2020hierarchical}.

\section{SIMULATION RESULTS}
To verify the efficiency of HPQ algorithm systematically and comprehensively, we conduct extensive simulations using SUMO \cite{lopez2018microscopic} to generate mixed random traffic flow. We select scenarios with traffic flows of approximately 1000 passenger car units/hour (pcu/h), 1300 pcu/h and 1600 pcu/h to simulate the algorithm and compare it with the hybrid AIM protocol proposed in \cite{2017A}, the delay-time actuated traffic lights algorithm proposed in \cite{oertel2011delay} and the distributed conflict-free cooperation algorithm proposed in \cite{XU2018322}.

The hybrid AIM protocol \cite{2017A} is a request-based right of way allocation algorithm for mixed traffic intersection management. Green trajectories and active green trajectories are defined to represent the right of way allocation of the intersection in real time. An active green trajectory is defined as a green trajectory with a HV present on it or on its incoming lane. If the trajectory of the CAV conflicts with the active green trajectory, the reservation of the CAV will be denied. Besides, the hybrid AIM protocol also investigates how to assign turning options for each lane and vehicle type and proposes turning assignment policy to improve traffic efficiency. The delay-time actuated traffic lights algorithm \cite{oertel2011delay} proposes an approach to control signal lamps by capturing the delay time of vehicles and utilize them to regulate the traffic signal timing. Within the range of the minimum and maximum green time, once the accumulated delay on a lane is dissolved, the running green phase is terminated. The distributed conflict-free cooperation algorithm \cite{XU2018322} is an iterative depth-first spanning tree algorithm. Each vehicle represents a vertice in the spanning tree. The depth of the vertices in the spanning tree is determined one by one according to the distance between the vehicle and the intersection center, and the conflict graph. According to the spanning tree, the right of way is granted to the corresponding vehicles. Further than these previous AIM algorithms, we comprehensively study the combination of the high-level HPQ algorithm and the low-level vehicle planning and control algorithm, and carry out detailed design for the vehicle control modes to improve traffic efficiency.

\begin{table}[!ht]

\renewcommand{\arraystretch}{1.5}
\caption{PARAMETER SETTINGS DURING THE CASE STUDY} \centering
\label{table_4}

\begin{tabular}{cm{3.7cm}c}
\toprule
\toprule
 Parameter  & Definition \centering &  Value \\
\hline
 $\sigma$  & The time constant of abnormal judgment \centering  & 100s \\
 $v_{limit}$    &  The speed limit for the road \centering  & $13.8m/s$\\
 $a_d$    & The maximum deceleration without loss of comfort \centering  & $-4m/s^2$ \cite{chen2020hierarchical}\\
 $a_{com}$    & The maximum acceleration without loss of comfort \centering  & $2m/s^2$  \cite{chen2020hierarchical}\\
 $l$   & The length of the entry area \centering  & $100m$\\
 $l_w$   & The width of the
lane \centering  & $3.5m$\\
 $L$  & The radius of the control range of the intersection area \centering & $103.5m$  \\
 $R$  & The radius of turning left \centering  & $5.25m$\\
 $r$   & The radius of turning right \centering   & $1.75m$\\
 $\epsilon$   & The distance constant  \centering & $20m$ \\
 $d_{min}$  & The minimum safety distance between the ego vehicle and the preceding vehicle \centering & \makecell[c]{$1m$ (CAVs) \cite{10021253}, \\ $2.5m$ (CHVs, HVs)}\\
 $h$  & The minimum travel time from the rear of the preceding vehicle to the front of the following vehicle \centering &  
\makecell[c]{$0.5s$ (CAVs) \cite{10021253}, \\ $2s$ (CHVs, HVs)}\\
$l_V$  & The length of
vehicle $V$ \centering & \
\makecell[c]{$5.2m$ (CAVs), \\ $4.0m$ (CHVs, HVs)}\\
\bottomrule
\bottomrule
\end{tabular}
\end{table}

Although SUMO can realize efficient mixed random traffic flow simulation in complex dynamic environment, it is arduous to integrate vehicle dynamics and verify the efficiency of the vehicle planning and control algorithms. Therefore, a joint simulation platform based on Matlab and PreScan \cite{tideman2010scenario} is built up to simulate the scenario of unsignalised intersections to verify the proposed vehicle planning and control algorithms. In the joint simulation platform based on Matlab and PreScan, Prescan can provide the realistic 3D modelling environment for traffic and physical models of sensor, and V2V and V2I communication simulation. Matlab can embed the vehicle planning and control algorithm to better realize algorithm simulation. The unsignalized intersection management strategy is validated by these simulation tools. The main fixed parameters used in the simulation are shown in Table III.


\subsection{Performance of HPQ Algorithm}
We use SUMO to generate mixed random traffic flow to evaluate the performance of the proposed HPQ algorithm. To ensure the safety of vehicles at intersections, the speed limit is 13.8 m/s. The simulation scenario of the HPQ algorithm is a four-way intersection covering 200 m $\times$ 200 m, as shown in Fig. 2. We use the average number of halts, average travel time and average speed to evaluate the performance of algorithms. Vehicle stop state refers to the state when the vehicle speed is less than 1.4 m/s. In order to better evaluate the performance of the HPQ algorithm under mixed autonomy traffic streams, the simulation is carried out when the CAV penetration rates are 50\%. The simulation results within 15 minutes are used for comparison.

\subsubsection{Traffic fluency} The average number of halts of vehicles under different traffic flows is calculated as shown in Fig. 10. It can be seen that both HPQ algorithm and the distributed conflict-free cooperation algorithm can greatly reduce the average number of halts of vehicles when the traffic flows are 1000 pcu/h and 1300 pcu/h. However, when the traffic flow increases to 1600 pcu/h, the HPQ algorithm is significantly better than the distributed conflict-free cooperation algorithm for traffic fluency. And the HPQ algorithm is always better than the delay-time actuated traffic lights algorithm and Hybrid AIM protocol under different traffic flows. Therefore, the HPQ algorithm can effectively reduce the number of halts of vehicles and greatly improve the traffic fluency at intersections.   


\begin{figure}[!t]
\centering
\includegraphics[width=3.5in]{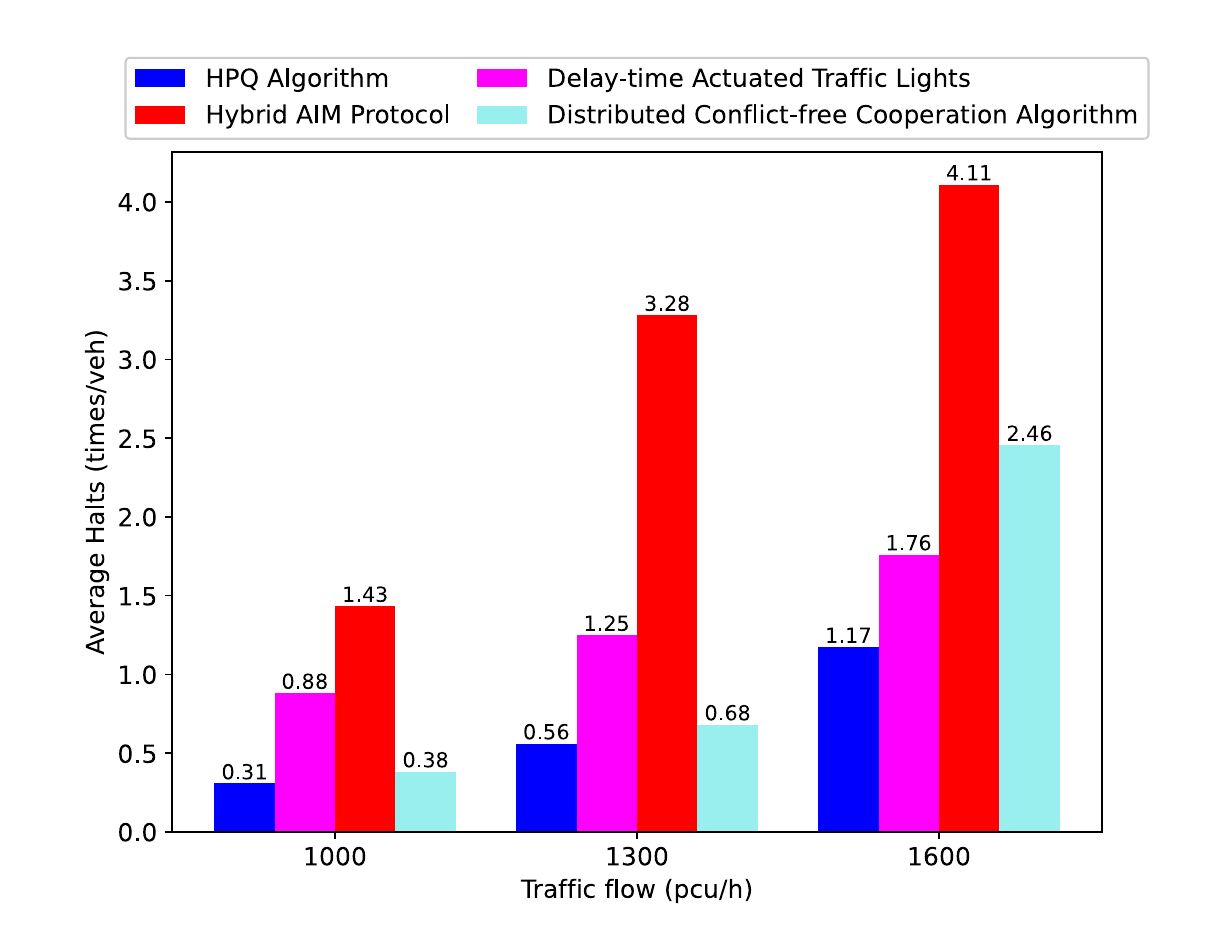}
\caption{Comparison of the number of halts of vehicles between different algorithms at the intersection under different traffic flows (within 15min).}
\label{fig_number_of_halts}
\end{figure}

\begin{figure*}[!t]
\centering
\subfloat[]{\includegraphics[width=2.35in]{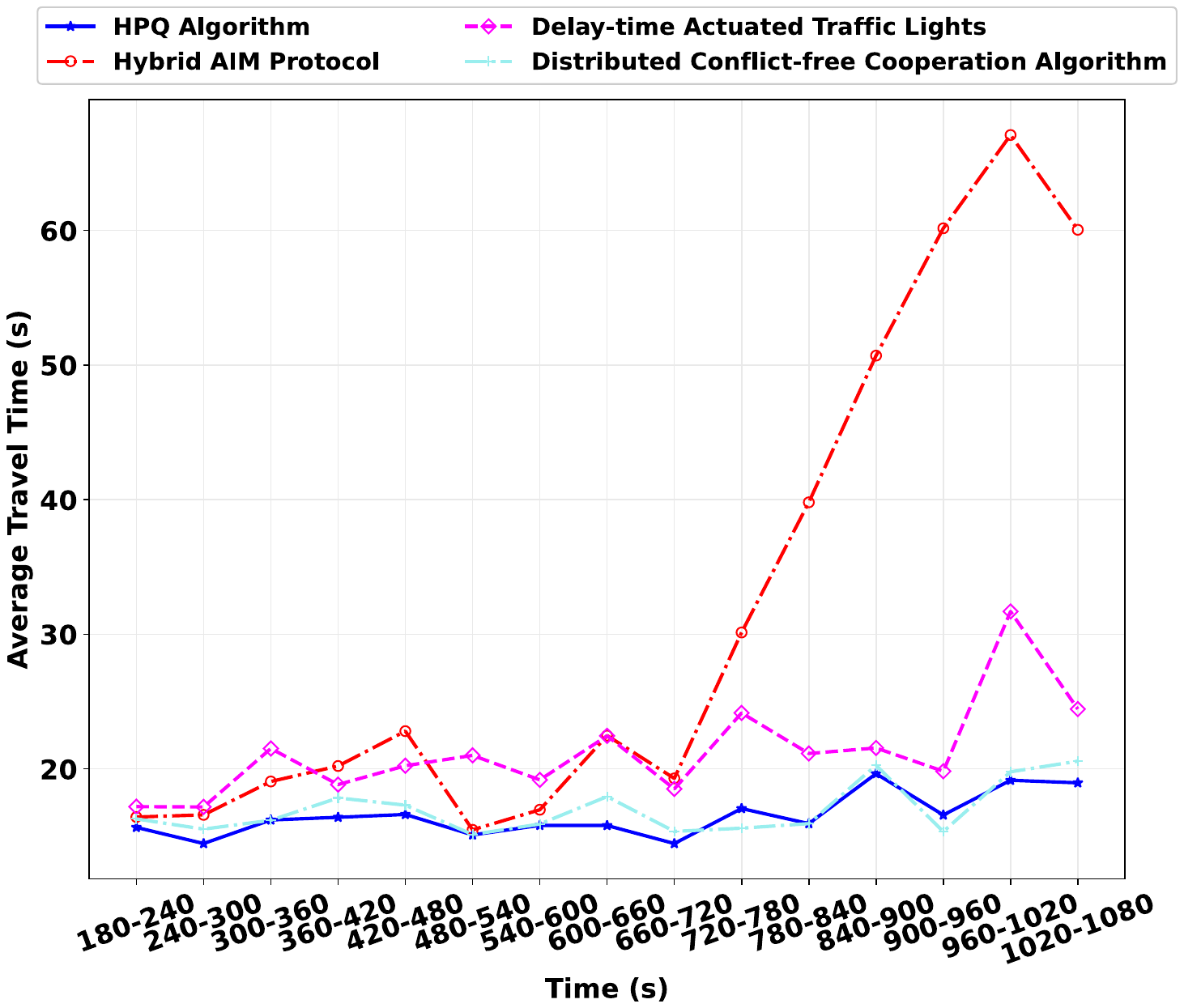}%
\label{fig_travel_time_first_case}}
\hfil
\subfloat[]{\includegraphics[width=2.35in]{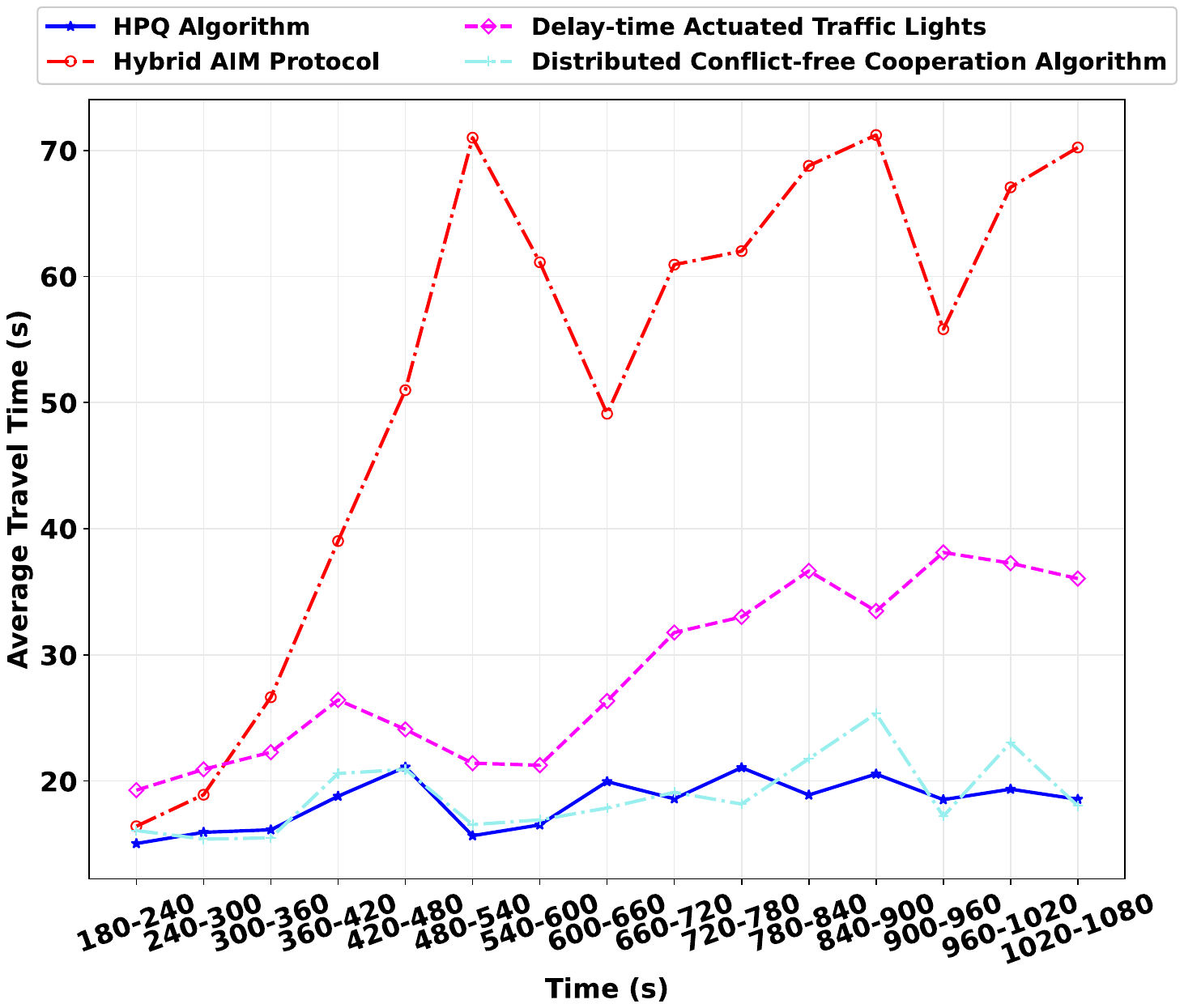}%
\label{fig_travel_time_second_case}}
\hfil
\subfloat[]{\includegraphics[width=2.35in]{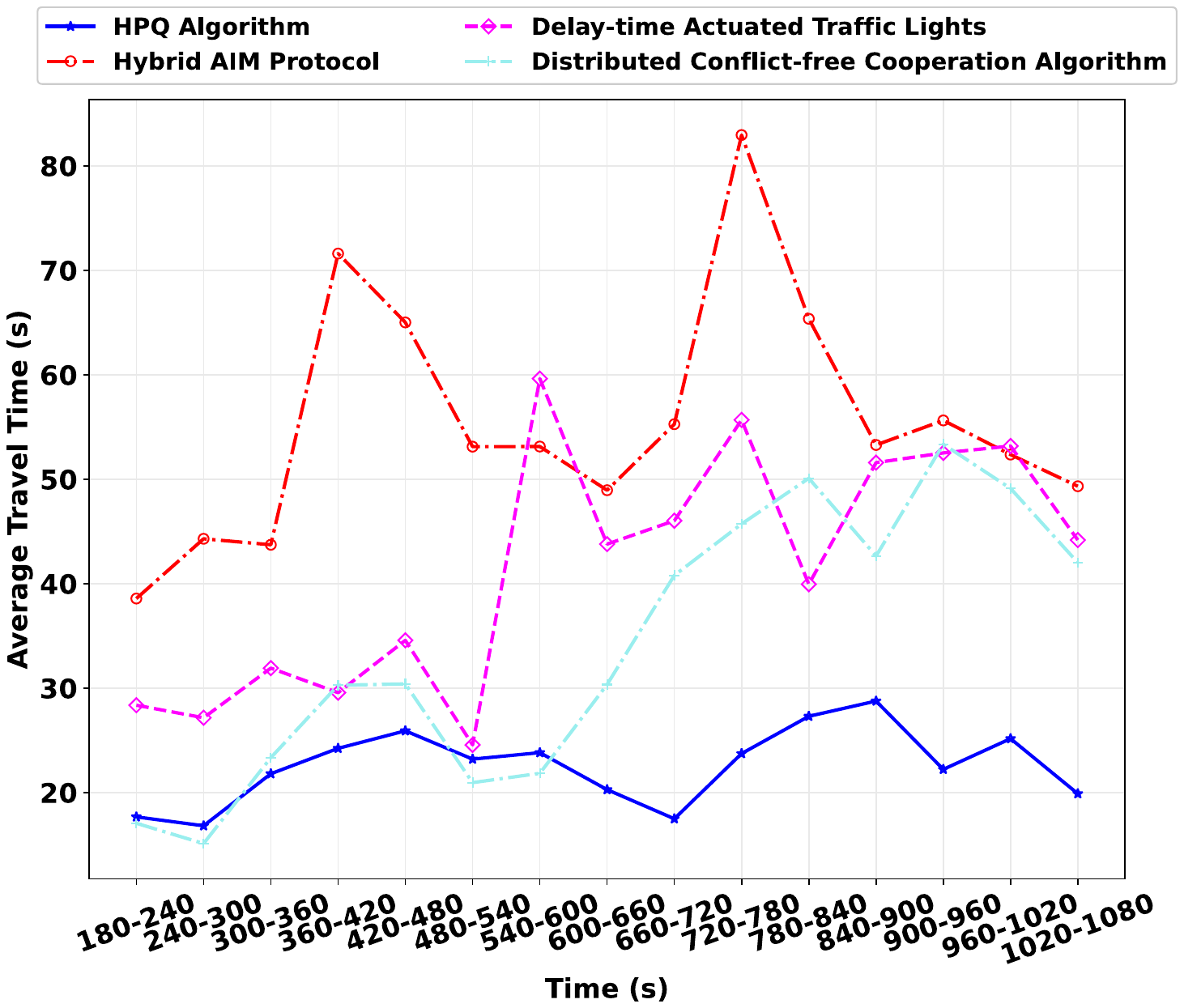}%
\label{fig_travel_time_third_case}}
\caption{Comparison of average travel time between different algorithms at the intersection under different traffic flows. (a) Light traffic (Traffic flow = 1000 pcu/h). (b) Moderate traffic (Traffic flow = 1300 pcu/h). (c) Heavy traffic (Traffic flow = 1600 pcu/h).}
\label{fig_travel_time_sim}
\end{figure*}

\begin{figure*}[!t]
\centering
\subfloat[]{\includegraphics[width=2.35in]{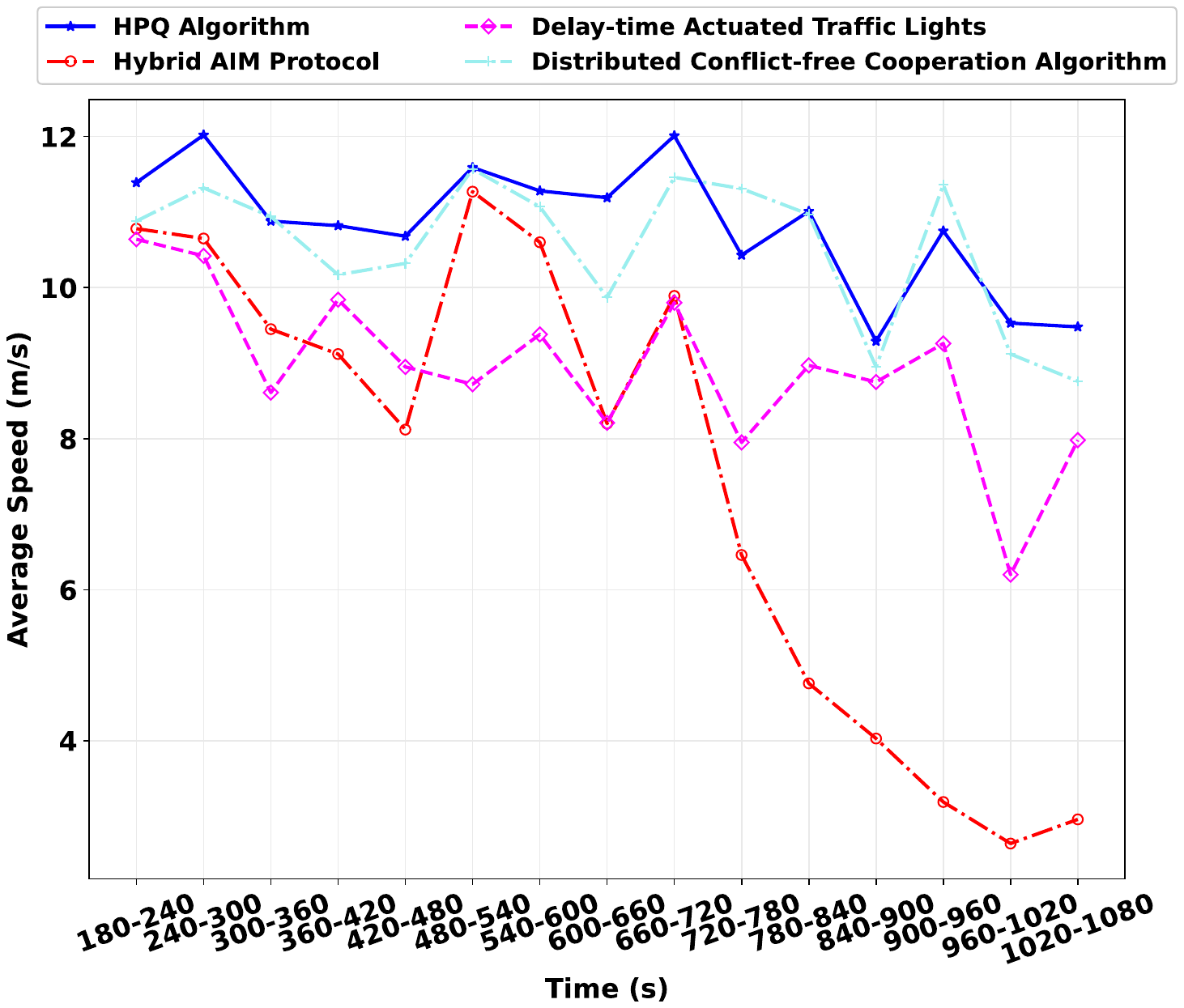}%
\label{fig_first_case}}
\hfil
\subfloat[]{\includegraphics[width=2.35in]{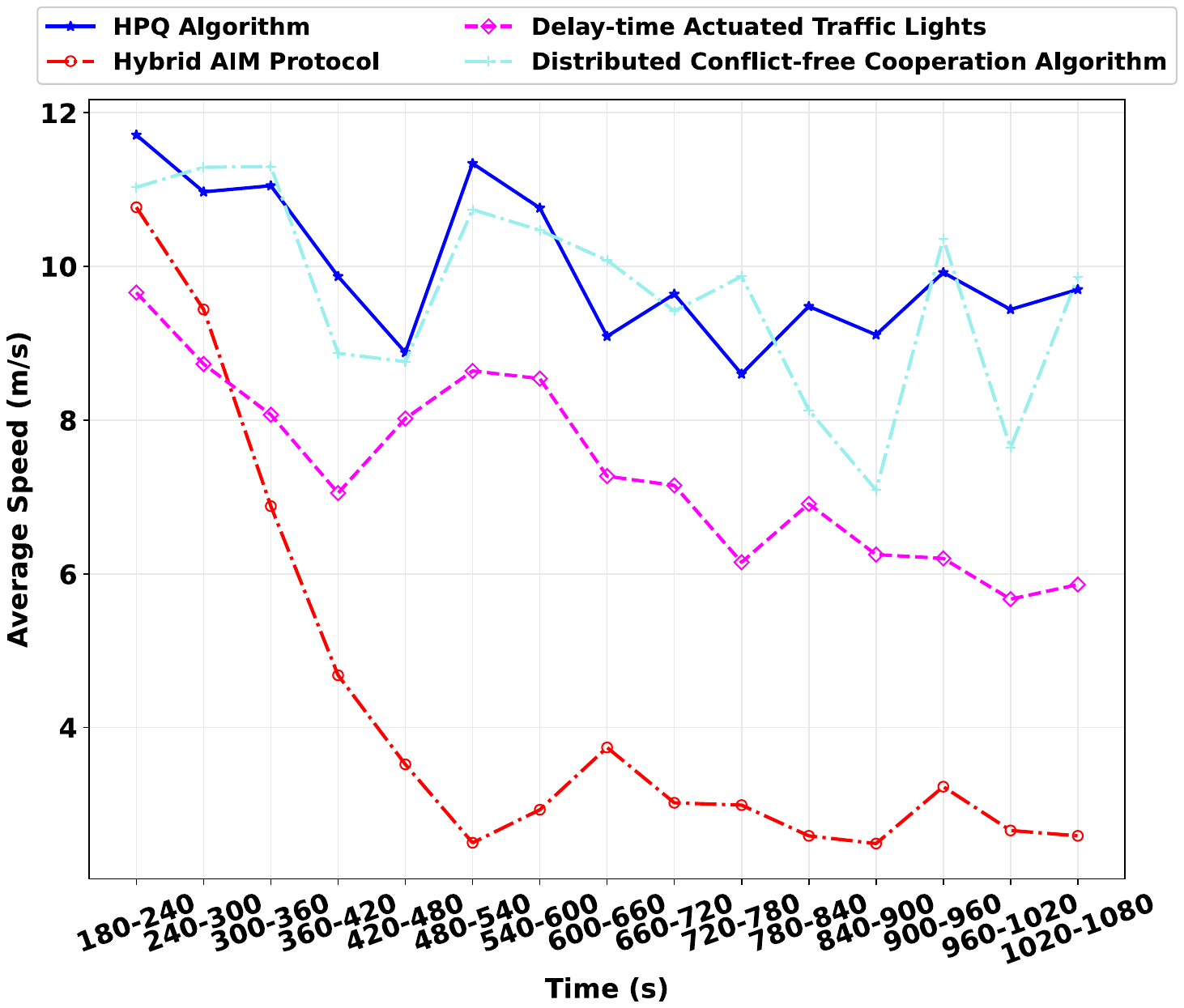}%
\label{fig_second_case}}
\hfil
\subfloat[]{\includegraphics[width=2.35in]{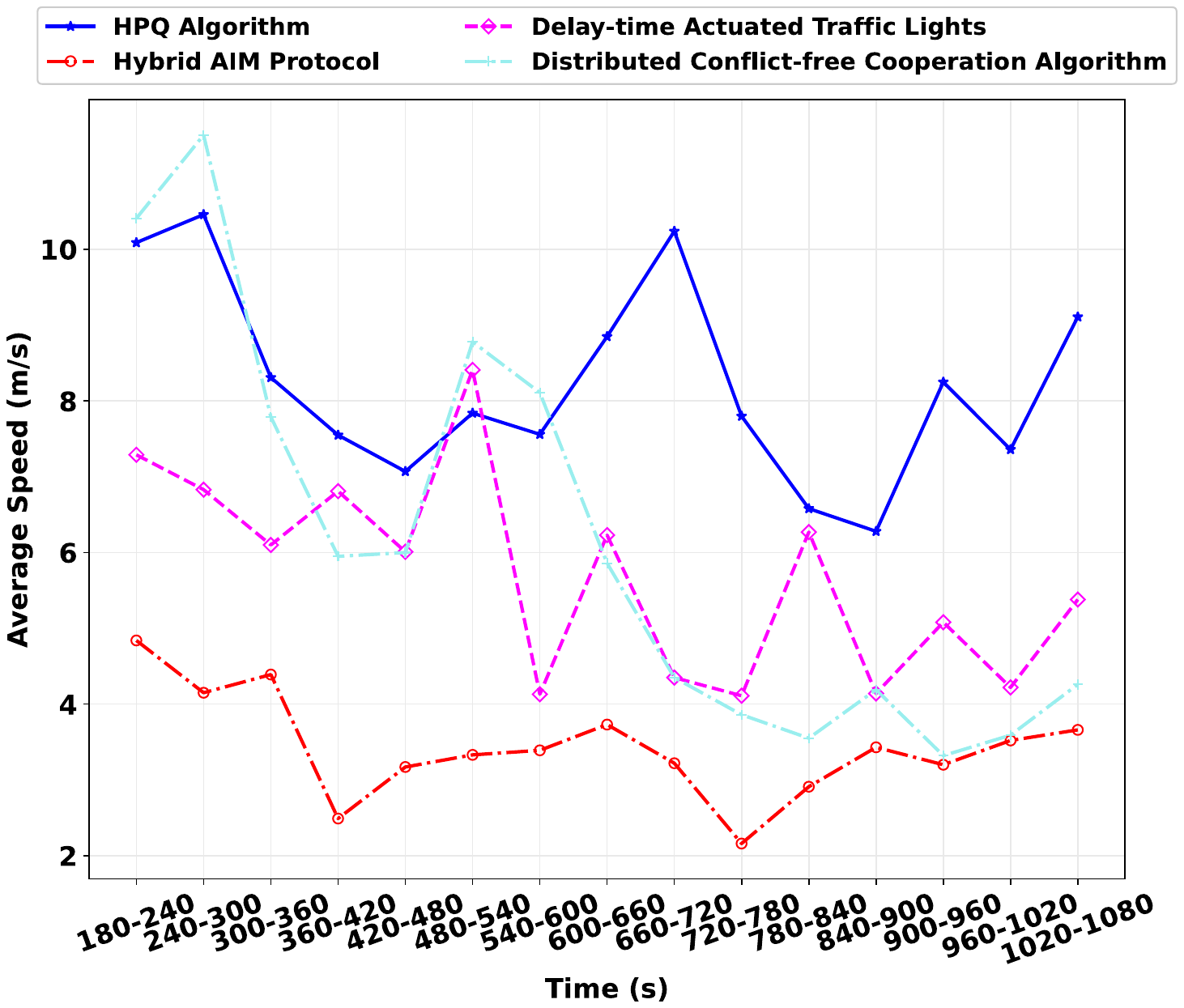}%
\label{fig_third_case}}
\caption{Comparison of average vehicle speed between different algorithms at the intersection under different traffic flows. (a) Light traffic (Traffic flow = 1000 pcu/h). (b) Moderate traffic (Traffic flow = 1300 pcu/h). (c) Heavy traffic (Traffic flow = 1600 pcu/h).}
\label{fig_speed_sim}
\end{figure*}

\subsubsection{Travel time} The travel time of vehicles under different traffic flows is shown in Fig. 11. It can be seen that both HPQ algorithm and the distributed conflict-free cooperation algorithm can greatly reduce the average travel time of vehicles when the traffic flows are 1000 pcu/h and 1300 pcu/h. All vehicles can pass through the intersections efficiently in a limited time. Moreover, the HPQ algorithm reduces the average travel time by up to 34\% compared to the distributed conflict-free cooperation algorithm when the traffic flow is 1600 pcu/h. And the distributed conflict-free cooperation algorithm is more easily impacted by traffic densities. Compared with the delay-time actuated traffic lights algorithm and Hybrid AIM protocol, the HPQ algorithm can reduce the average travel time of vehicles at intersections by 22\% to 65\% under different traffic flows. The traffic light scheme seriously restrains traffic efficiency of intersections. In particular, the travel times of the HPQ algorithm at different periods and different traffic flows are very stable, comparing with the fluctuations of the other three algorithms. Therefore, the HPQ algorithm can effectively reduce the travel time of vehicles at intersections, so that vehicles can pass through the intersections more smoothly.

\subsubsection{Average speed} To further investigate the effectiveness of the HPQ algorithm, we evaluate their performance using average vehicle speed, as shown in Fig. 12. It can be seen that vehicles under the control of the HPQ algorithm can pass through the intersection smoothly at a relatively high speed under different traffic flows. However, the speed of vehicles under the control of the other three algorithms fluctuates widely. This can be well explained since the vehicles under the control of the other three algorithms encounter a greater number of halts. The HPQ algorithm can reduce the number of halts and increase the speed of vehicles at intersections, therefore effectively improve the comfort and safety of occupants.


In general, the vehicles at the intersection under the control of the HPQ algorithm have a significant improvement in the traffic efficiency compared with the vehicles at the intersection under the control of the delay-time actuated traffic lights algorithm, the Hybrid AIM protocol and the distributed conflict-free cooperation algorithm. In addition, the performance of the HPQ algorithm is stable with different situations. Under low density traffic streams, vehicles almost do not need to slow down and stop at the intersection to avoid collision. 

\subsection{Simulations of Vehicle Planning and Control Algorithm}
Fig. 13 shows the comprehensive simulation scenario configuration. Table IV illustrates the parameters used in simulation for vehicle planning and control algorithm based on intersection coordinate system. According to the HPQ algorithm and the vehicle planning and control algorithm proposed in this paper, the control modes of all vehicles are as follows:

\begin{figure}[!t]
\centering
\includegraphics[width=3.1in]{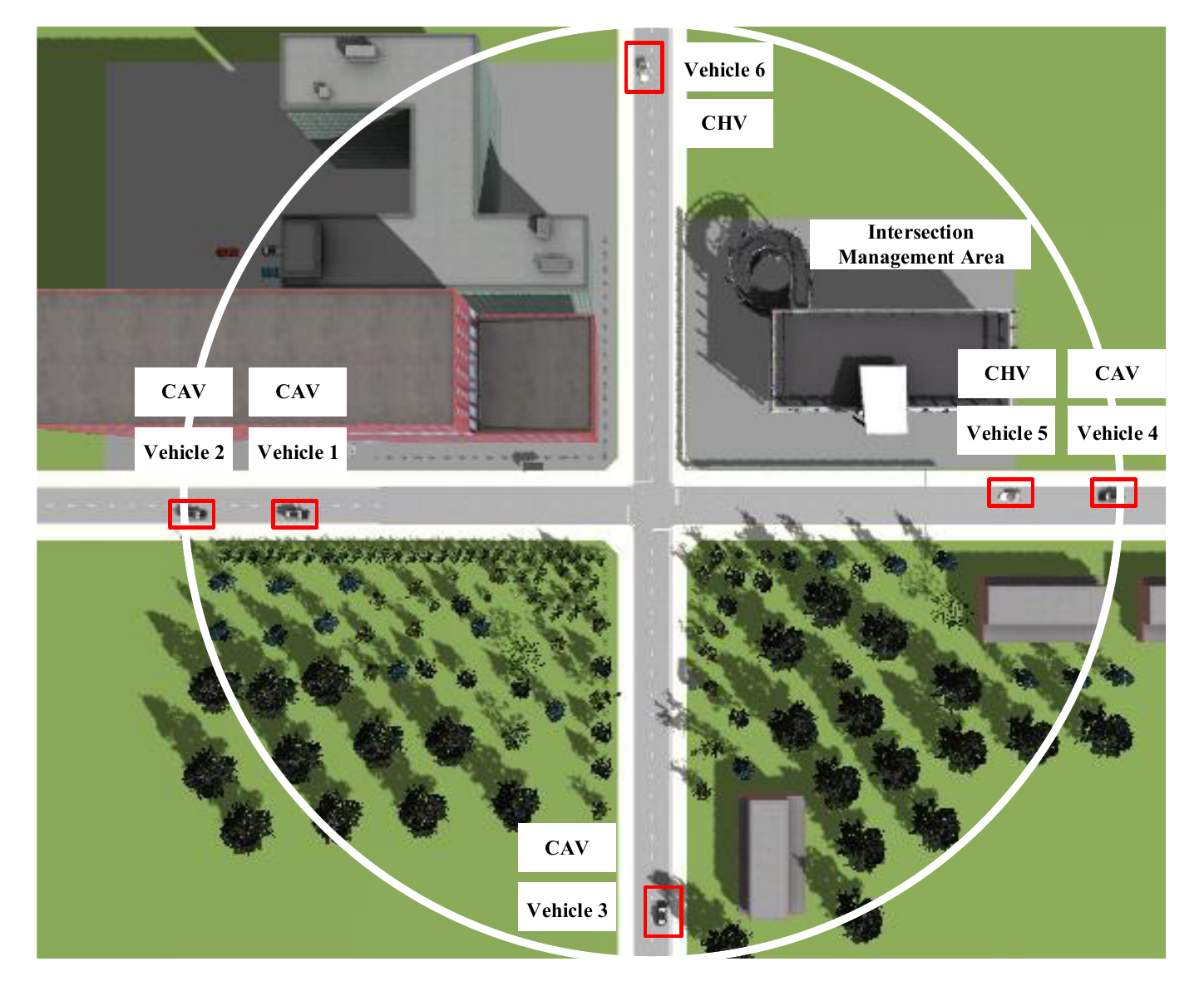}
\caption{Comprehensive simulation scenario configuration.}
\label{fig_PreScan}
\end{figure}

\begin{table}[!t]

\renewcommand{\arraystretch}{1.3}
\caption{PARAMETERS USED IN SIMULATION FOR COMPREHENSIVE SCENARIO}
\label{table_4}
\centering
\begin{tabular}{|c|c|c|c|c|c|}
\hline
\makecell[c]{Vehicle\\ number} & \makecell[c]{Vehicle\\ type} & Priority & \makecell[c]{Driving\\ direction} & \makecell[c]{Initial\\ position} & \makecell[c]{Initial\\ speed}\\ 
\hline
1 & CAV & 2 & Turn left & (-70 m, -1.75 m) & 9 m/s\\
\hline
2 & CAV & 6 & Straight & (-90 m, -1.75 m) & 9 m/s\\
\hline
3 & CAV & 3 & Straight & (1.75 m, -80 m) & 9 m/s\\
\hline
4 & CAV & 5 & Straight & (90 m, 1.75 m) & 9 m/s\\
\hline
5 & CHV & 1 & Turn right & (70 m, 1.75 m) & 9 m/s\\
\hline
6 & CHV & 4 & Turn left & (-1.75 m, 85 m) & 9 m/s\\
\hline
\end{tabular}
\end{table}

\begin{figure}[!t]
\centering
\includegraphics[width=3.5in]{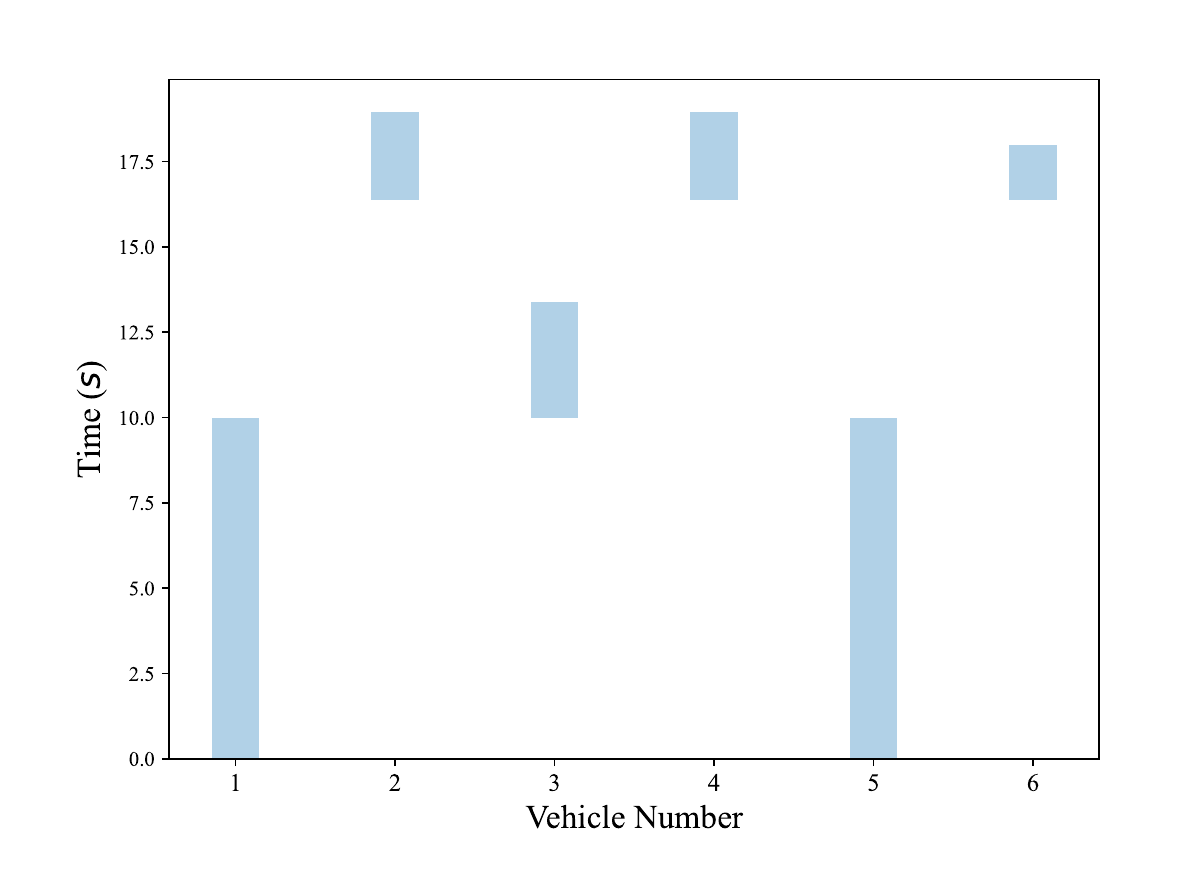}
\caption{Right of way allocation in comprehensive scenario.}
\label{fig_right_of_way}
\end{figure}

\begin{figure}[!t]
\centering
\includegraphics[width=3.5in]{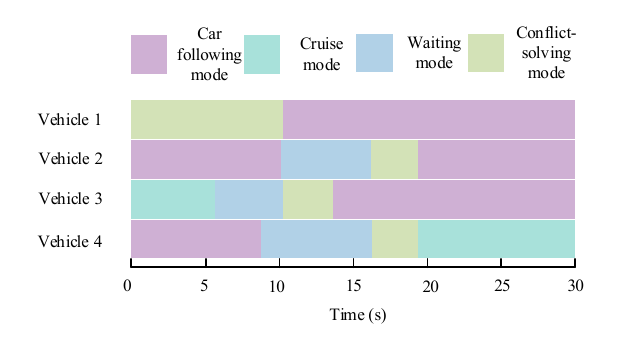}
\caption{Control modes switching of CAVs in the comprehensive scenario.}
\label{fig_control_mode}
\end{figure}



\begin{figure}[!t]
\centering
\subfloat[]{\includegraphics[width=3.5in]{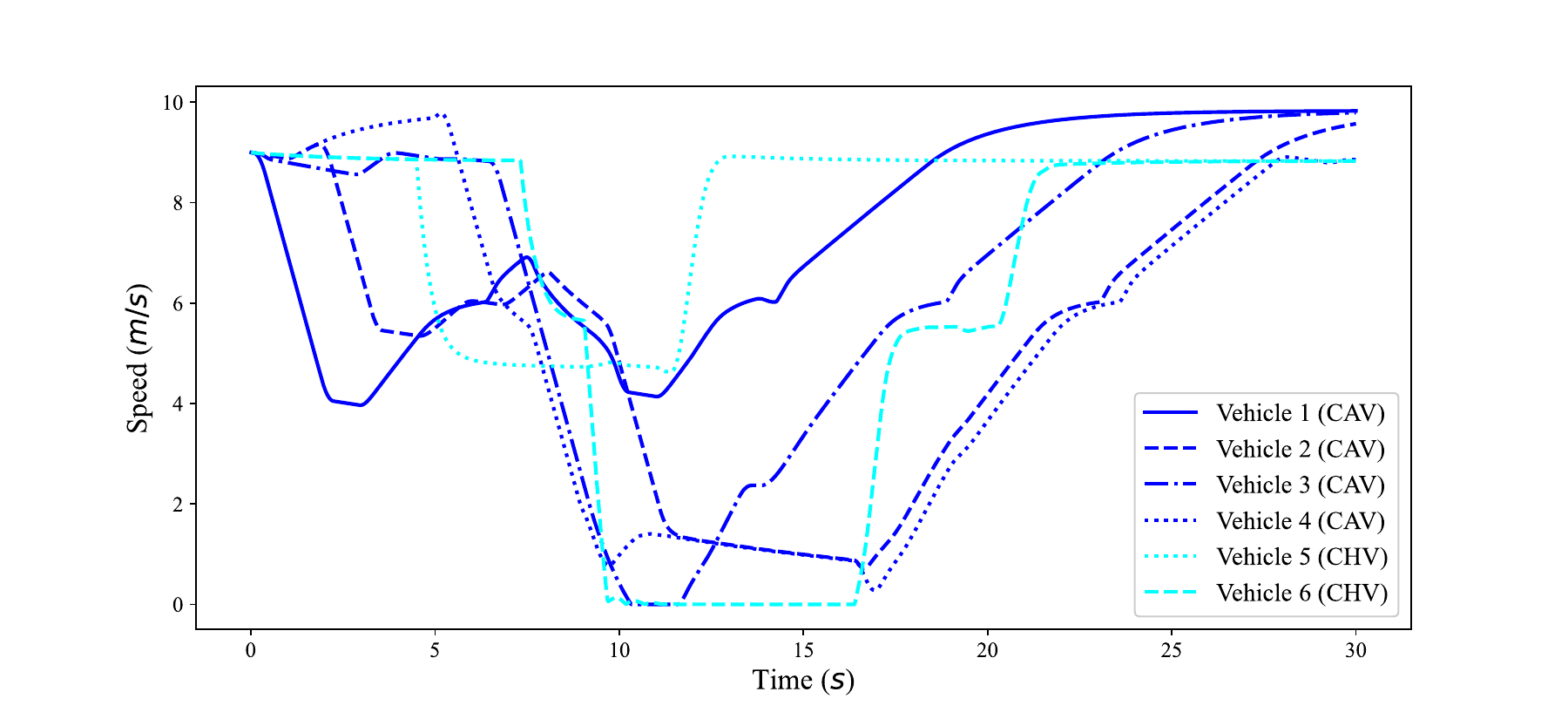}%
\label{fig_vehicle_distance1}}
\hfil
\subfloat[]{\includegraphics[width=3.5in]{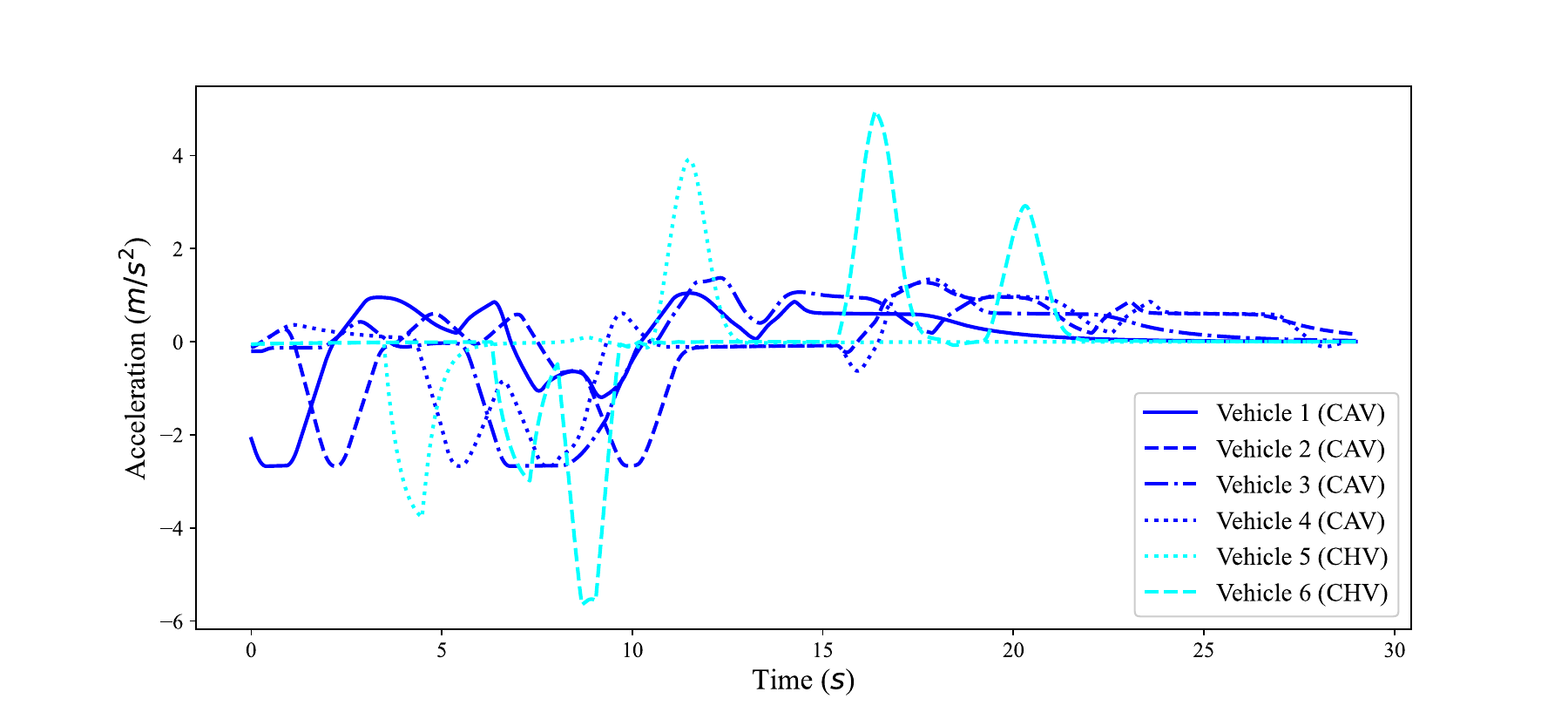}%
\label{fig_vehicle_speed1}}
\caption{Speed profiles and acceleration profiles of CAVs and CHVs in the comprehensive scenario. (a) Speed profiles of CAVs and CHVs in the comprehensive scenario, where the blue line indicates the speed profile of the CAV and the cyan line indicates the speed profile of the CHV. (b) Acceleration profiles of CAVs and CHVs in the comprehensive scenario.}
\label{fig_accel and speed of vehicle 3}
\end{figure}

\begin{figure}[!t]
\centering
\subfloat[]{\includegraphics[width=3.5in]{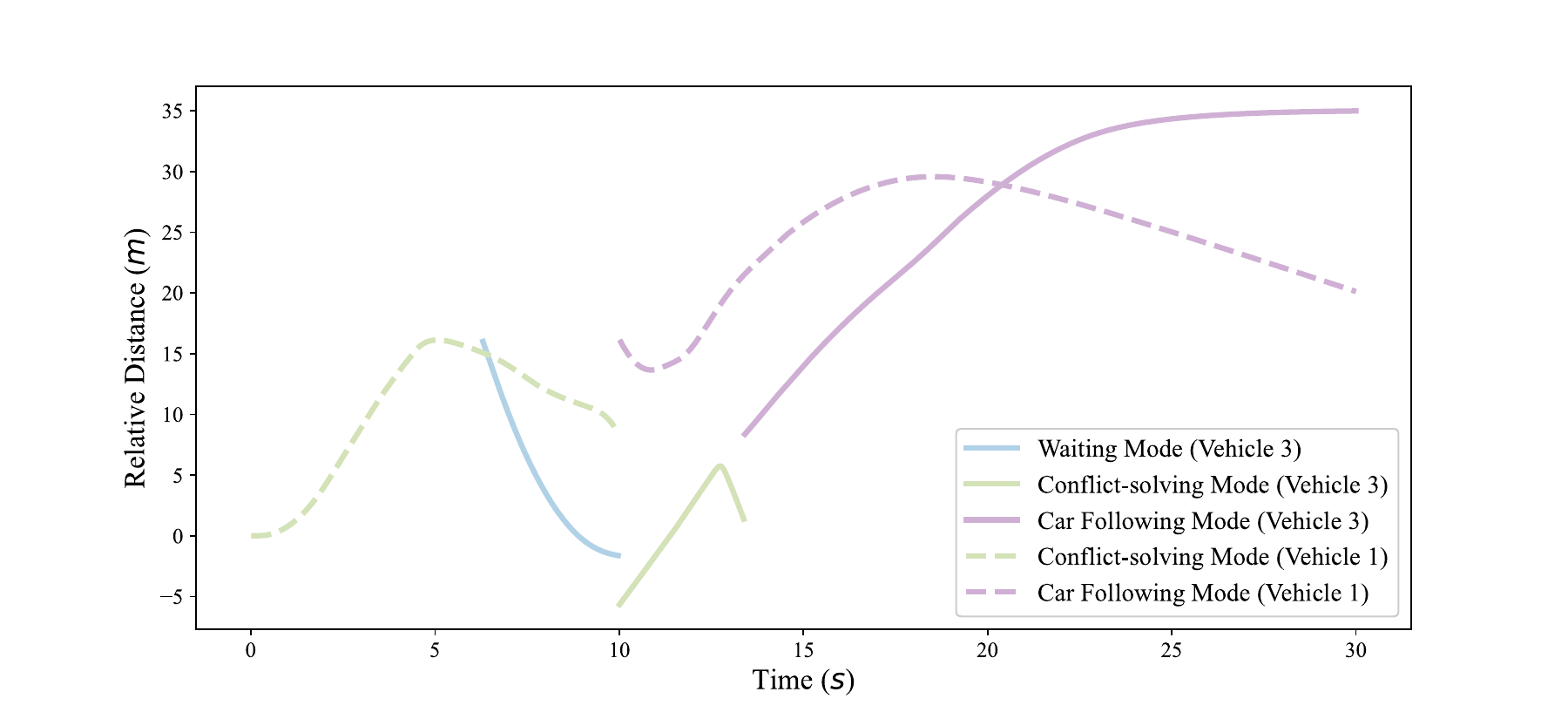}%
\label{fig_vehicle_distance1}}
\hfil
\subfloat[]{\includegraphics[width=3.5in]{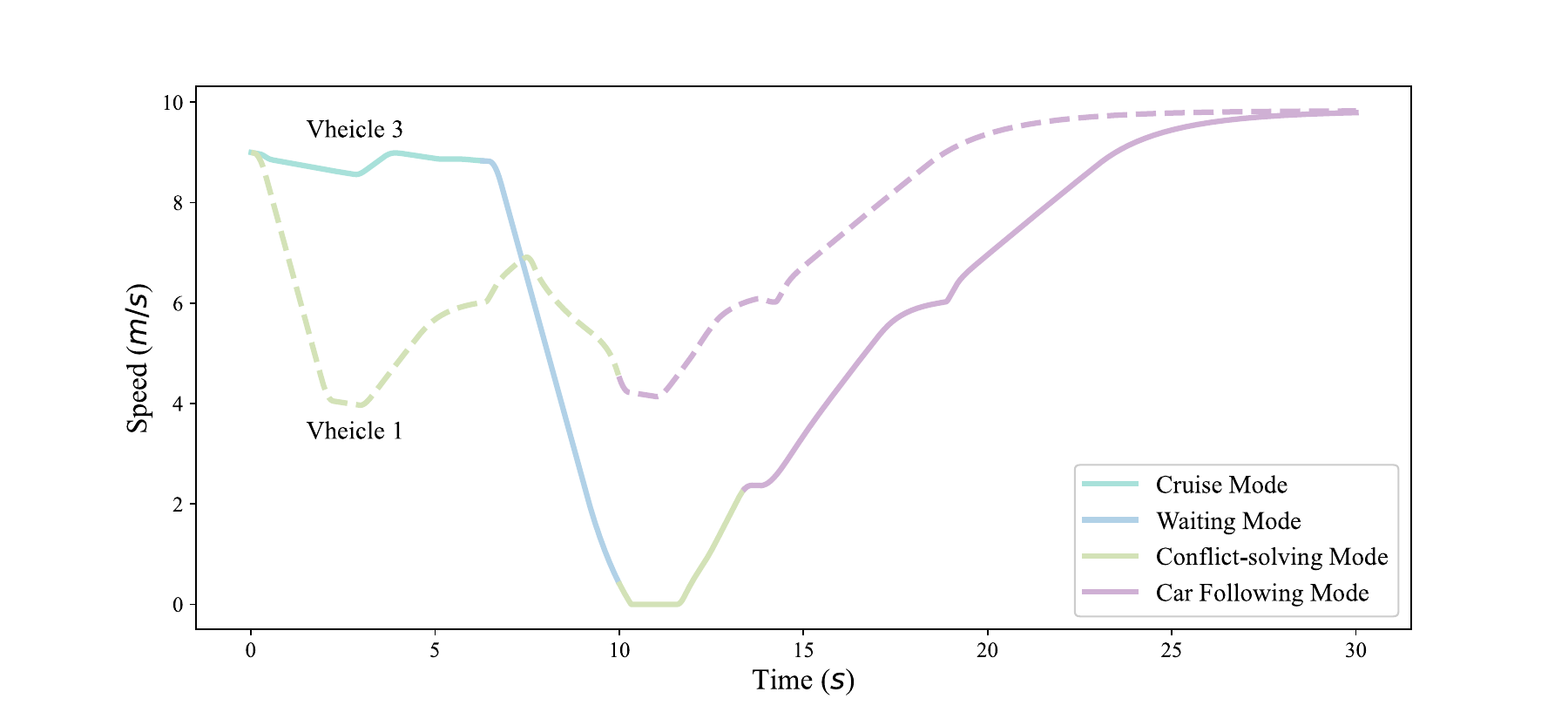}%
\label{fig_vehicle_speed1}}
\caption{Relative distance and speed of vehicle 1 and vehicle 3. (a) Relative distances of vehicle 1 and vehicle 3 to the surrounding reference vehicles in different control modes. (b) Speed of vehicle 1 and vehicle 3 in different control modes.}
\label{fig_Relative distance and speed of vehicle 3}
\end{figure}

\begin{itemize}
\item Vehicle 1: Firstly, vehicle 1 obtains the right of way and switches to the conflict-solving mode to coordinate the merging conflict with vehicle 5. After vehicle 5 passed through the conflict area, vehicle 1 switches to car following mode to follow vehicle 5. 
\item Vehicle 2: Vehicle 2 follows vehicle 1 when entering the intersection control area. When vehicle 2 gets near the stop line, vehicle 2 switches to the waiting mode because vehicle 2 cannot get the right of way. It is not until vehicle 3 passed through the conflict area that vehicle 2 obtains the right of way and switches to the conflict-solving mode to coordinate the merging conflict with vehicle 6. After vehicle 6 passed through the conflict area, vehicle 2 switches to car following mode to follow vehicle 6.
\item Vehicle 3: When vehicle 3 enters the intersection control area, it first switches to cruise mode. When vehicle 3 is near the stop line, it will switch to the waiting mode because it has not obtained the right of way. Vehicle 3 will not get the right of way until the vehicle 5 passes the conflict area, and then vehicle 3 switches to the conflict-solving mode to coordinate the merging conflict with the vehicle 1. After vehicle 1 passed through the conflict area, vehicle 3 switches to car following mode to follow vehicle 1.
\item Vehicle 4: When vehicle 4 enters the intersection control area, it follows vehicle 5 first. when vehicle 4 approaches the stop line, it switches to waiting mode because it has not obtained the right of way. vehicle 4 will obtain the right of way and switch to the conflict-solving mode to coordinate the crossing conflict with vehicle 6 when vehicle 3 passed through the conflict area. After vehicle 6 passed through the conflict area, vehicle 4 switches to cruise mode.
\item Vehicle 5: When vehicle 5 arrives at the intersection control area, it will obtain the right of way, and the driver will drive at a constant speed.
\item Vehicle 6: The driver first drives at a constant speed to the stop line and waits. Until vehicle 3 passes through the conflict area, vehicle 6 will obtain the right of way and continue to drive at a constant speed through the intersection.
\end{itemize}

During the whole simulation, the allocation of the right of way at the intersection is shown in Fig. 14. The blue bars represent the time when the vehicle is granted the right of way. We can see that vehicle 1 and vehicle 5 obtain the right of way first. The control modes switching of CAVs are shown in Fig. 15. Different colors represent different control modes of CAVs. When the color of the bar is blue, it means that the vehicle is in car following mode. We can see that vehicle 2 and vehicle 4 are in car following mode at the beginning of the simulation.


The speed profiles and acceleration profiles of CAVs and CHVs in the comprehensive simulation scenario are illustrated in Fig. 16. Fig. 16(a) depicts the speed profiles of CAVs and CHVs in the comprehensive scenario, where the blue line indicates the speed profile of the CAV and the cyan line indicates the speed profile of the CHV. Fig. 16(b) depicts the acceleration profiles of CAVs and CHVs in the comprehensive scenario. Vehicles 1-4 represent the CAVs, and vehicles 5-6 represent the CHVs. It can be seen that the speed profiles of the CAVs are smooth enough. In addition, the acceleration of the CAVs is basically kept between $-3m/s^2$ and $2m/s^2$, which not only ensures the safety of the vehicles, but also improves the comfort of the occupants. In contrast, the speed profile of the CHVs is slightly less smooth, and its maximum acceleration is greater than $4m/s^2$. The main reason for this is that all the CAVs near the intersection are controlled by the designed lower-level vehicle planning and control algorithm, which combines the upper-level HPQ algorithm to calculate the control modes for each CAV and uses model predictive control to control the vehicles safely, quickly and comfortably through the intersection. The car following model for the CHVs uses the classical Krauss model \cite{krauss1997metastable}, under the assumption that humans can visually detect the distance to the vehicle ahead and drive safely to maintain a desired distance from it \cite{bifulco2022decentralized}. Thus the CAVs can achieve smooth motor behavior and improved driving comfort based on the idea of rolling optimization in model predictive control, whereas the behaviors of the CHVs are prone to more fluctuations influenced by the surrounding environment.

The relative distance and speed of vehicle 1 and vehicle 3 are shown in Fig. 17. Fig. 17(a) shows the relative distance of vehicle 1 and vehicle 3 to the surrounding reference vehicles, and Fig. 17(b) shows the speed of vehicle 1 and vehicle 3 in different control modes. It can be seen from the figure that vehicle 1 and vehicle 3 can quickly and accurately switch to the corresponding control mode designed under different traffic situations. And vehicle 1 and vehicle 3 can maintain a safe distance from surrounding vehicles at a relatively stable speed. Undisturbed switching of four control modes not only guarantees vehicle safety, but also improves the utilization rate of the conflict area of the intersection.

\begin{figure}[!t]
\centering
\subfloat[]{\includegraphics[width=1.58in]{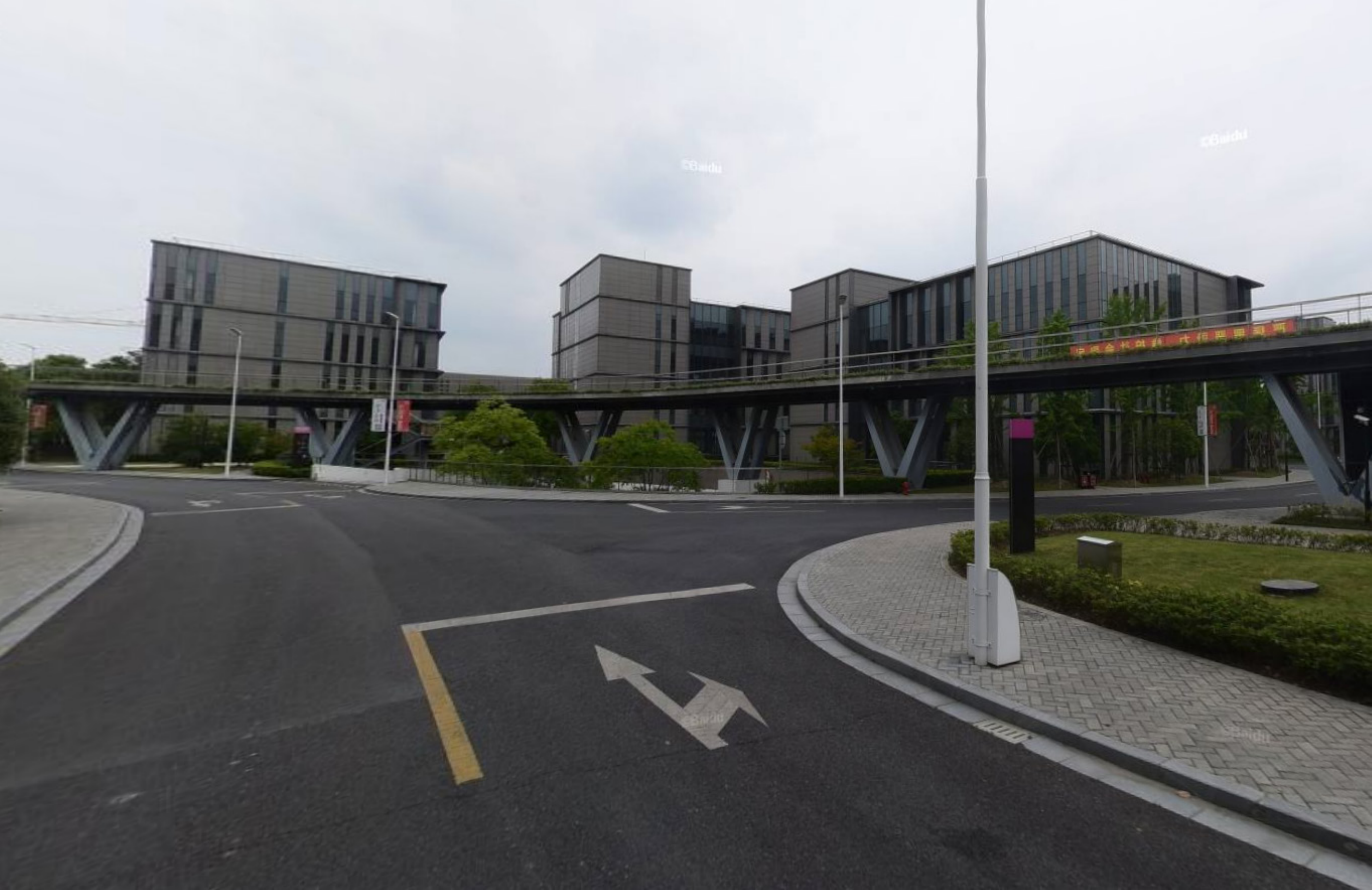}%
\label{fig_intersection modeling}}
\hfil
\subfloat[]{\includegraphics[width=1.58in]{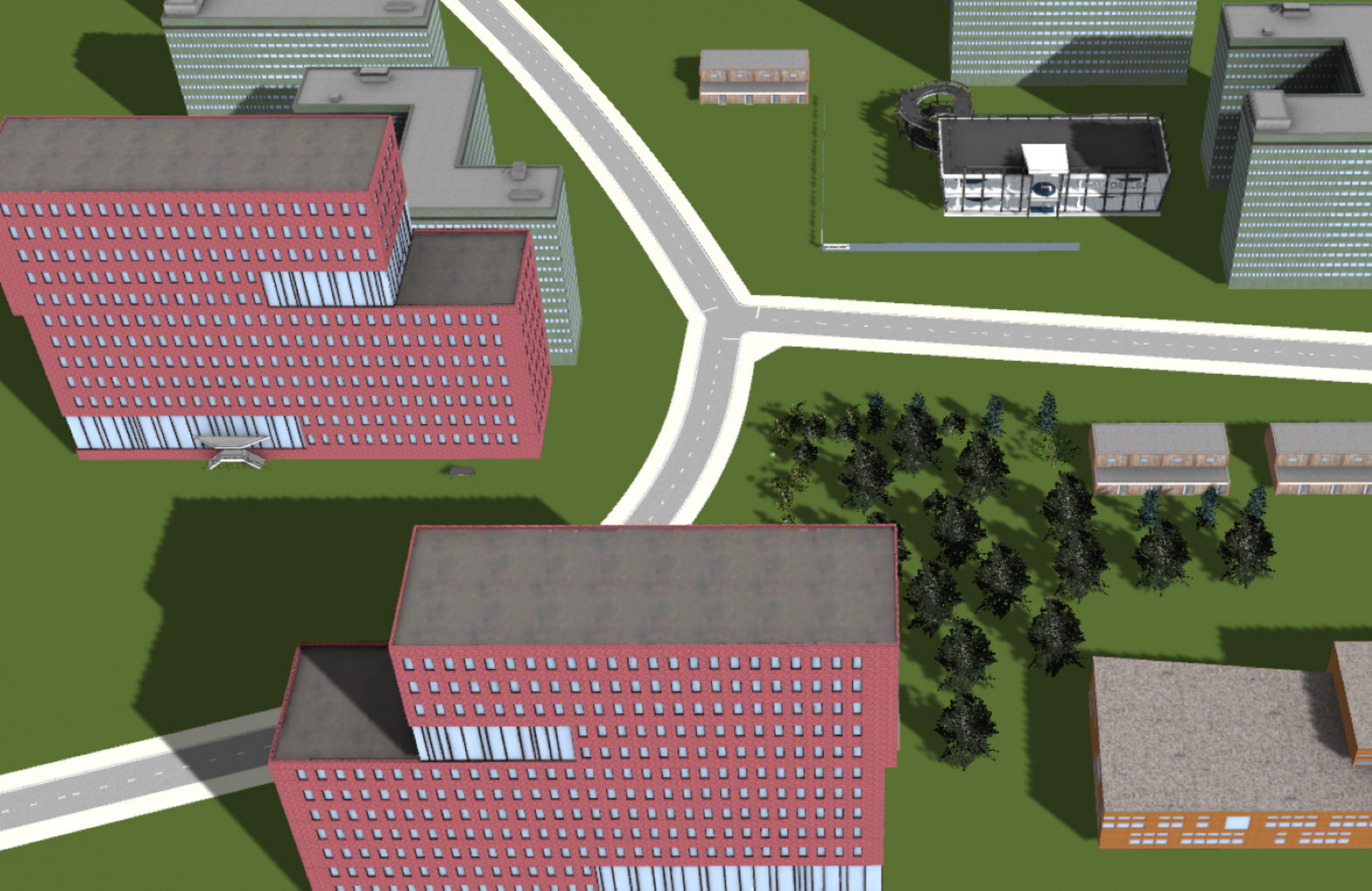}%
\label{fig_intersection modeling2}}
\caption{The scenario layout of the intersection. (a) The actual photo of the simulated intersection. (b) The simulation environment built in PreScan according to the real scene.}
\label{fig_intersection_modeling}
\end{figure}

To better verify the effectiveness and robustness of the proposed unsignalized intersection management strategy, we conduct a larger-scale simulation considering the intersection at Shanghai Intelligent and Connected Vehicle Test Zone. The scenario layout of the intersection is shown in Fig. 18, where Fig. 18(a) is the actual photo of the simulated intersection, and Fig. 18(b) is the simulation environment built in PreScan according to the real scene in Fig. 18(a). The vehicles arrive at the intersection entrances randomly with the initial speed of 9 m/s. The right of way of the vehicles is calculated by the high-level HPQ algorithm. Then, the high-level calculation results are transmitted to the low-level vehicle terminal for vehicle planning and control. There are a total of 30 vehicles participated in the simulation.

Fig. 19 presents the arrival time of different types of vehicles approaching the intersection in different traffic movements. The circular contours indicate the different levels of arrival time, and their values are displayed in the color bar of the figure. Besides, the arrows indicate different vehicle movements. It can be seen from Fig. 18 that there are three entry areas at the intersection, and each entry area is connected to two exit areas. The solid dots and hollow dots in Fig. 19 represent the time when the CAVs and the CHVs arrive at the intersection, respectively. The points on the same circle correspond to vehicles passing through the intersection at the same time. It can be seen from the figure that there is no conflict between vehicles passing through the intersection at the same time and all vehicles can pass through the intersection within 60 s, which proves that the unsignalized intersection management strategy can coordinate the vehicles passing the intersection safely and efficiently. In addition, the computation load required by the unsignalized intersection management strategy is small enough, so it is highly feasible to be applied to actual traffic control scenarios. 

\begin{figure}[!t]
\centering
\includegraphics[width=3.5in]{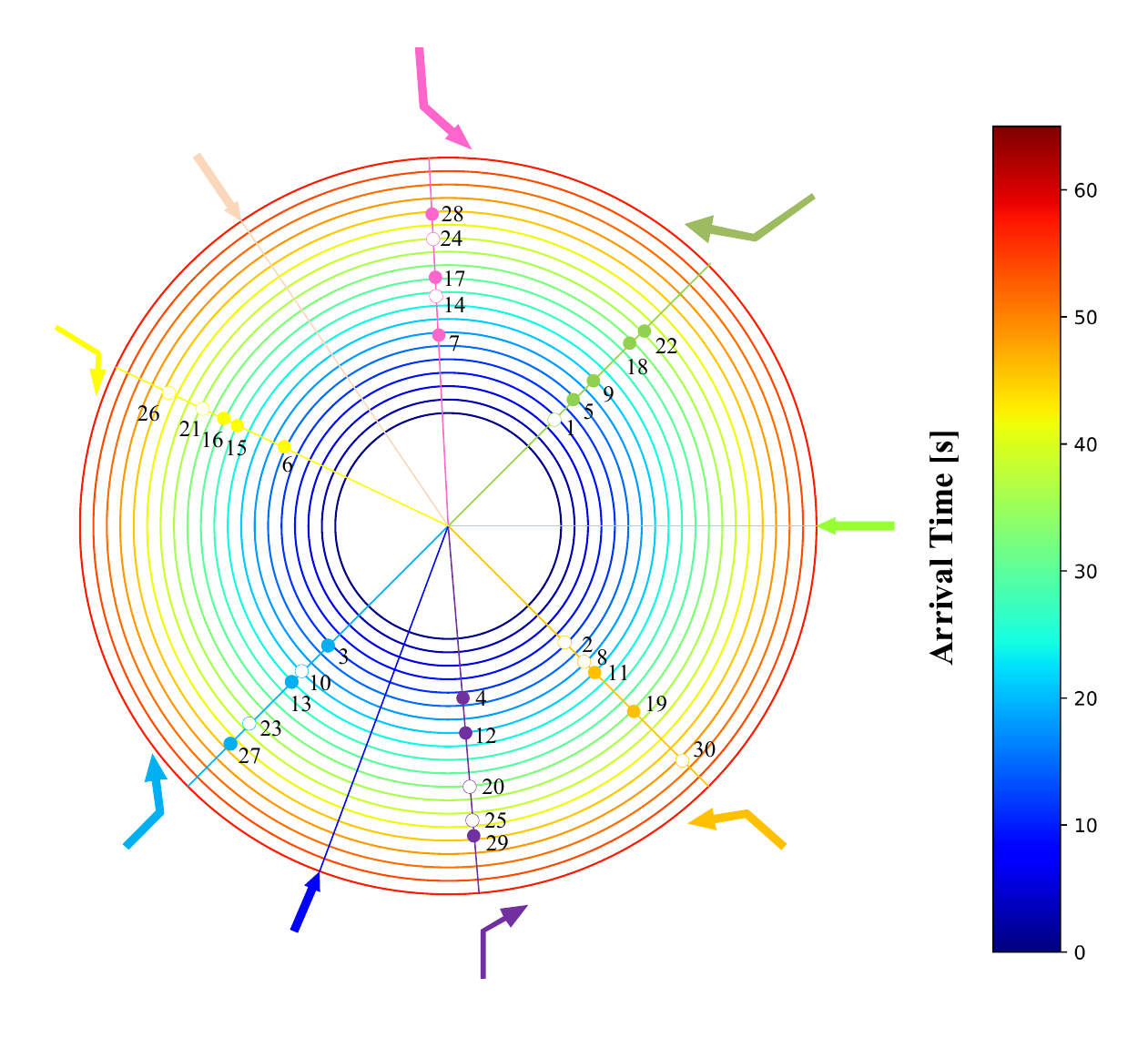}
\caption{Arrival time of approaching vehicles at three-way, one-lane irregular intersection simulations.}
\label{fig_Arrival time}
\end{figure}

\subsection{Micro-simulation Validation of Mixed Autonomy Traffic Streams}

To demonstrate the effectiveness of the proposed method in mixed autonomy traffic streams, we conducted micro-simulations across a broader range of intersection configurations and traffic scenarios. Specifically, we constructed a digital twin simulation scenario in PreScan micro-simulation software based on the real-world four-way, three-lane intersection to demonstrate that the proposed method can be applied to many applications and problem sizes. The scenario layout of the intersection is illustrated in Fig. 20, where Fig. 20(a) is the actual photo of the case study at the intersection between Jianchuan Road and Dushi Road near Shanghai Jiao Tong University, and Fig. 20(b) is the simulation environment built in
PreScan according to the real scene in Fig. 20(a). In the four-way, three-lane intersection, we conducted a comprehensive simulation experiment of mixed traffic flow, including CAVs, CHVs, and HVs.

\begin{figure}[!t]
\centering
\subfloat[]{\includegraphics[width=1.6in]{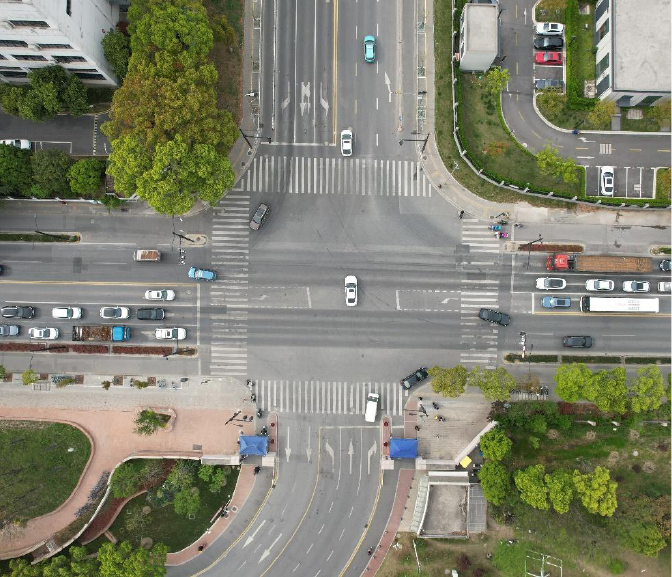}%
\label{fig_intersection modeling}}
\hfil
\subfloat[]{\includegraphics[width=1.6in]{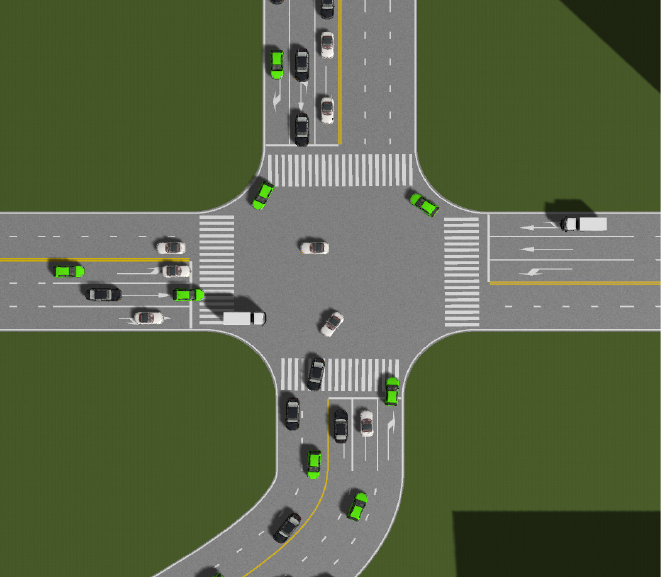}%
\label{fig_intersection modeling2}}
\caption{One of the most congested intersections near Shanghai Jiao Tong University is depicted in 20(a), and the simulation environment built in PreScan, according to the real scene, is illustrated in 20(b).}
\label{fig_intersection_modeling}
\end{figure}

\begin{figure}[!t]
\centering
\includegraphics[width=3.5in]{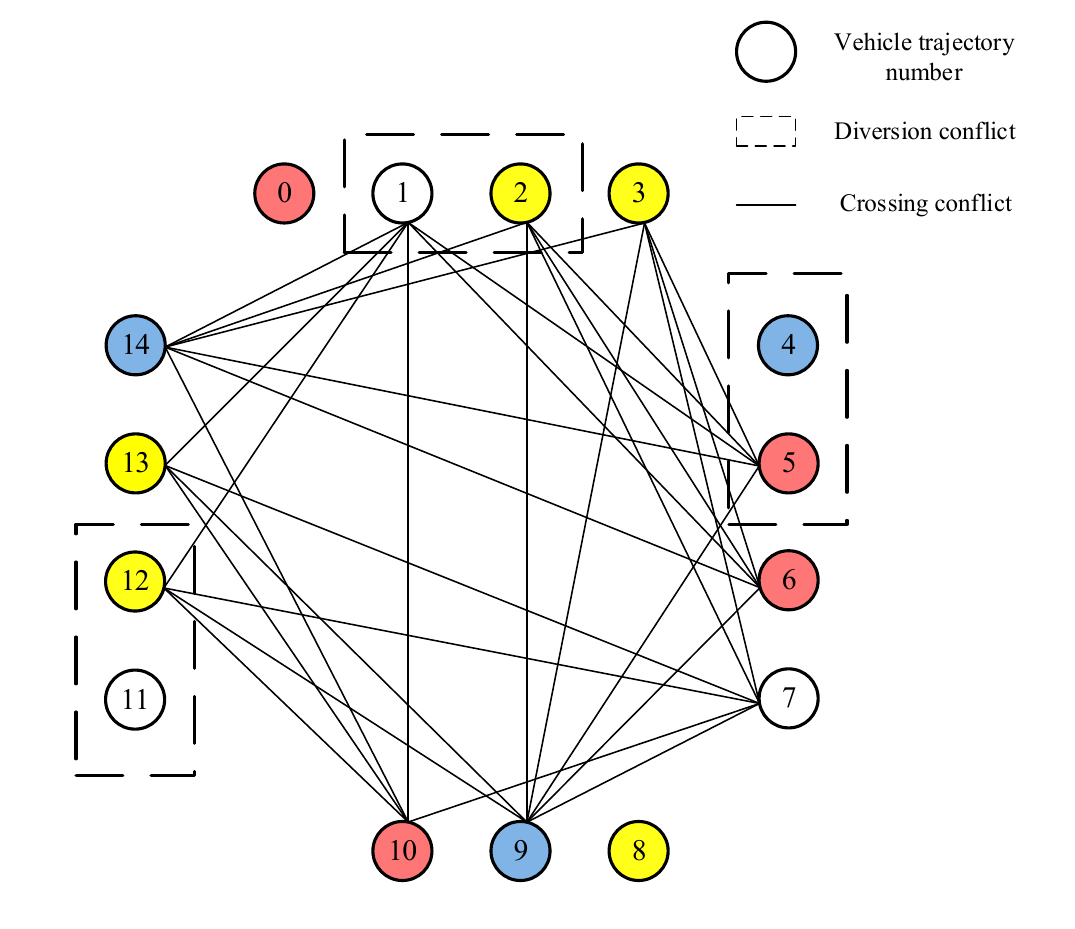}
\caption{Conflict relationship of traffic movements at the four-way, three-lane intersection in the comprehensive simulation scenario under mixed traffic streams. Circles of the same color indicate merging conflicts between trajectories. Dashed outlines denote diversion conflicts between trajectories within the dashed box. Intersection points represent crossing conflicts between crossing trajectories.}
\label{fig_Vehicle_coordinate_system}
\end{figure}

We use PreScan to validate the proposed unsignalized intersection management method for large-scale intersection scenarios under mixed traffic flows in microsimulation. The relationship diagram of vehicle motion conflicts in the simulation is shown in Fig. 21. From Fig. 21, it can be seen that there are three types of vehicle conflicts in this scenario: diversion conflict, crossing conflict, and merging conflict. The profile of the distance between the position of the vehicle and the center of the intersection is shown in Fig. 22. Different types of vehicles are represented by different line types. Vehicles 1-14 are CAVs. Vehicles 15-24 are CHVs. Vehicles 25-38 are HVs. The top dotted line indicates the stop line at the intersection boundary, and the bottom dotted line indicates the center of the intersection. It can be seen from Fig. 22 that most vehicles have a smooth trajectory and can safely pass through the intersection in an orderly manner. Different types of vehicles can be effectively coordinated to pass through the intersection, and the CAVs reserve a certain safety distance for the HVs to ensure that the HVs can pass through the intersection safely. This demonstrates the robustness and effectiveness of the proposed method.



\begin{figure}[!t]
\centering
\includegraphics[width=3.5in]{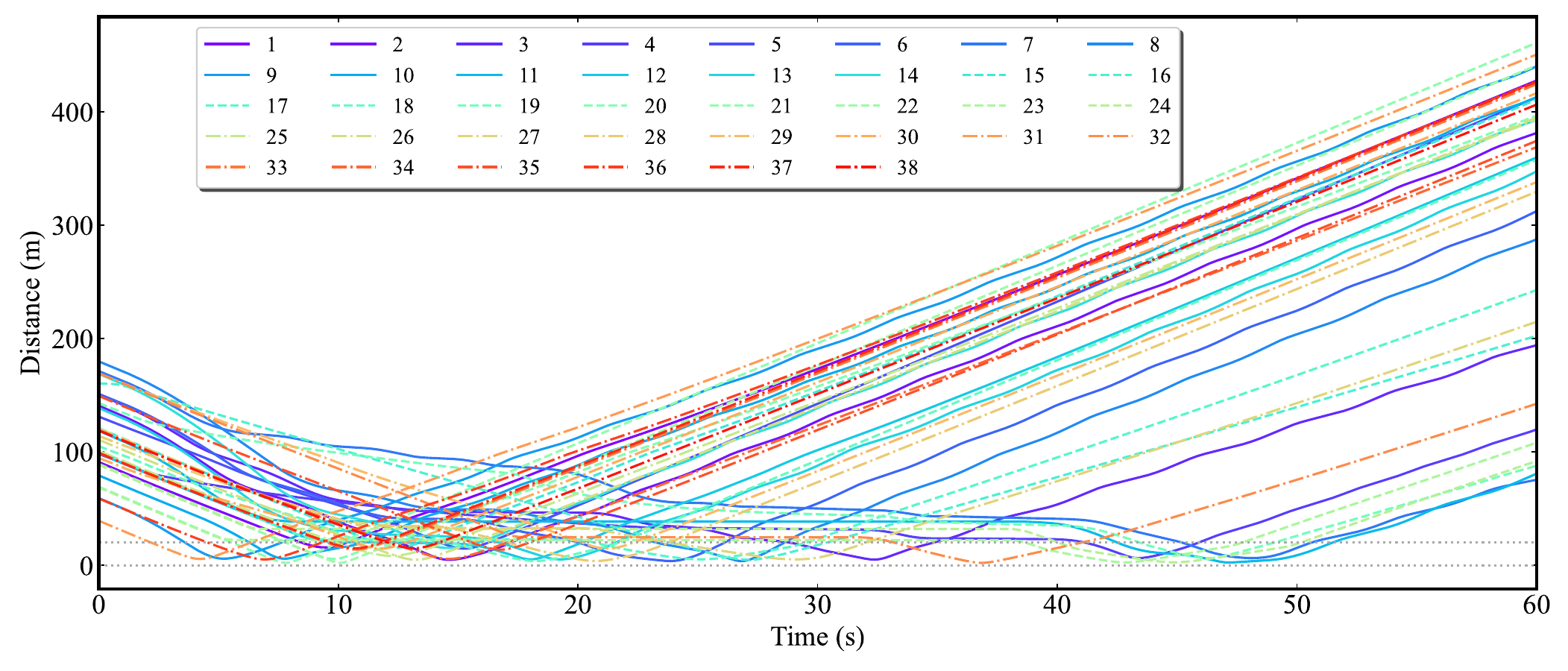}
\caption{Profile of the distance between the vehicle's position and the intersection's center. Different types of vehicles are represented by different line types. Vehicles 1-14 are CAVs. Vehicles 15-24 are CHVs. Vehicles 25-38 are HVs. The top dotted line indicates the stop line at the intersection boundary, and the bottom dotted line indicates the center of the intersection.}
\label{fig_Vehicle_coordinate_system}
\end{figure}

\begin{figure}[!t]
\centering
\includegraphics[width=3.5in]{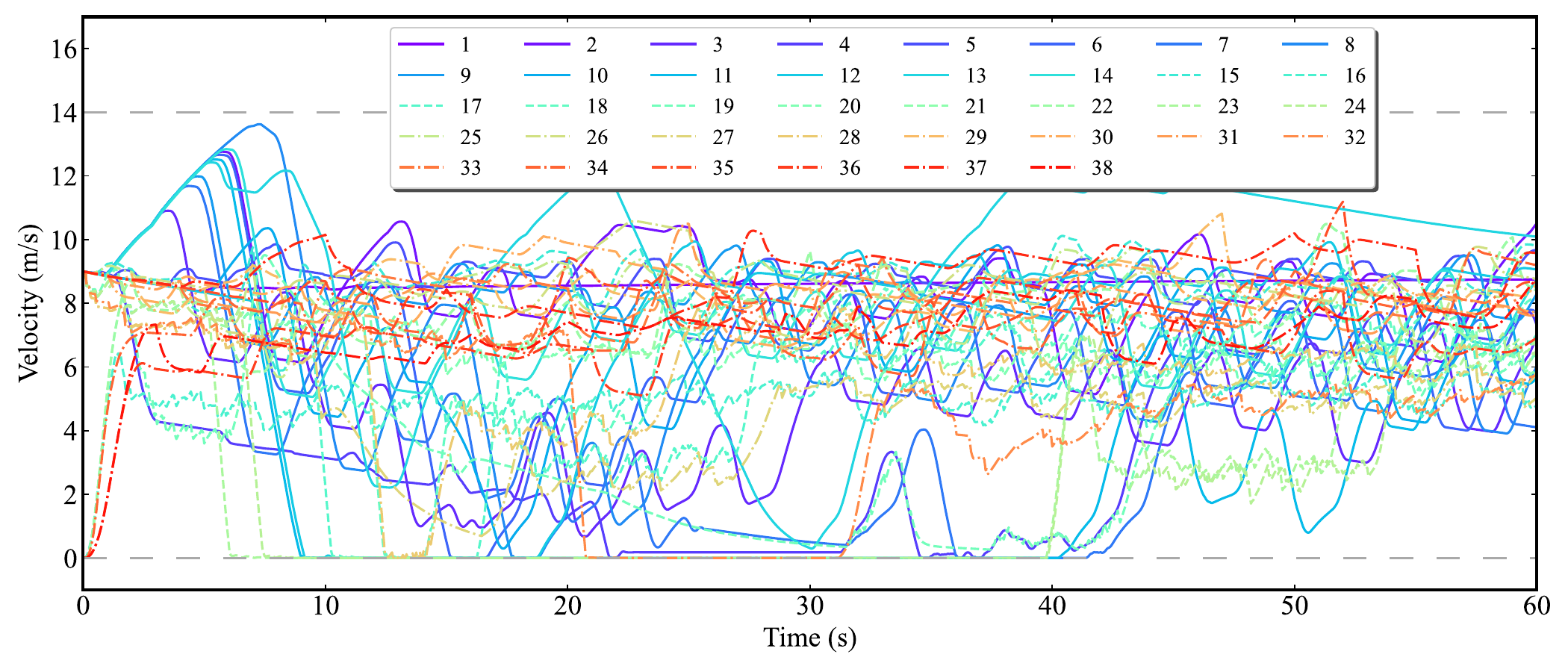}
\caption{Vehicle velocity profiles in the complex intersection under mixed traffic flow. The speed profiles of CAVs exhibit smoother behavior in contrast to that of HVs, which aligns with real-world driving scenarios. Most of the vehicles pass through the intersection quickly at high speeds. Notably, CAVs proactively yield to the HVs in close proximity to the intersection to mitigate trajectory conflicts.}
\label{fig_Vehicle_coordinate_system}
\end{figure}

\begin{figure*}[!t]
\centering
\includegraphics[width=7in]{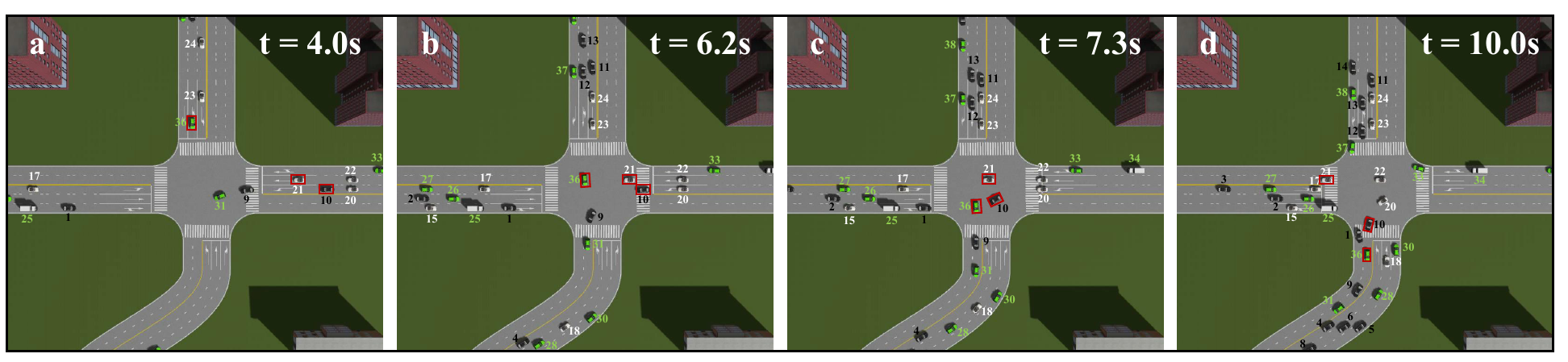}
\caption{Snapshots of the typical conflict scenario solved by the proposed unsignalized intersection management method under the mixed traffic streams of CAVs, CHVs, and HVs. Vehicles 1-14 are CAVs. Vehicles 15-24 are CHVs. Vehicles 25-38 are HVs. The red rectangles in the snapshot highlight critical moments in the mixed traffic flow of CAVs, CHVs, and HVs. There is a typical conflict relationship between CAV 10, CHV 21, and HV 36.}
\label{fig_Vehicle_coordinate_system}
\end{figure*}

\begin{figure}[!t]
\centering
\includegraphics[width=3.5in]{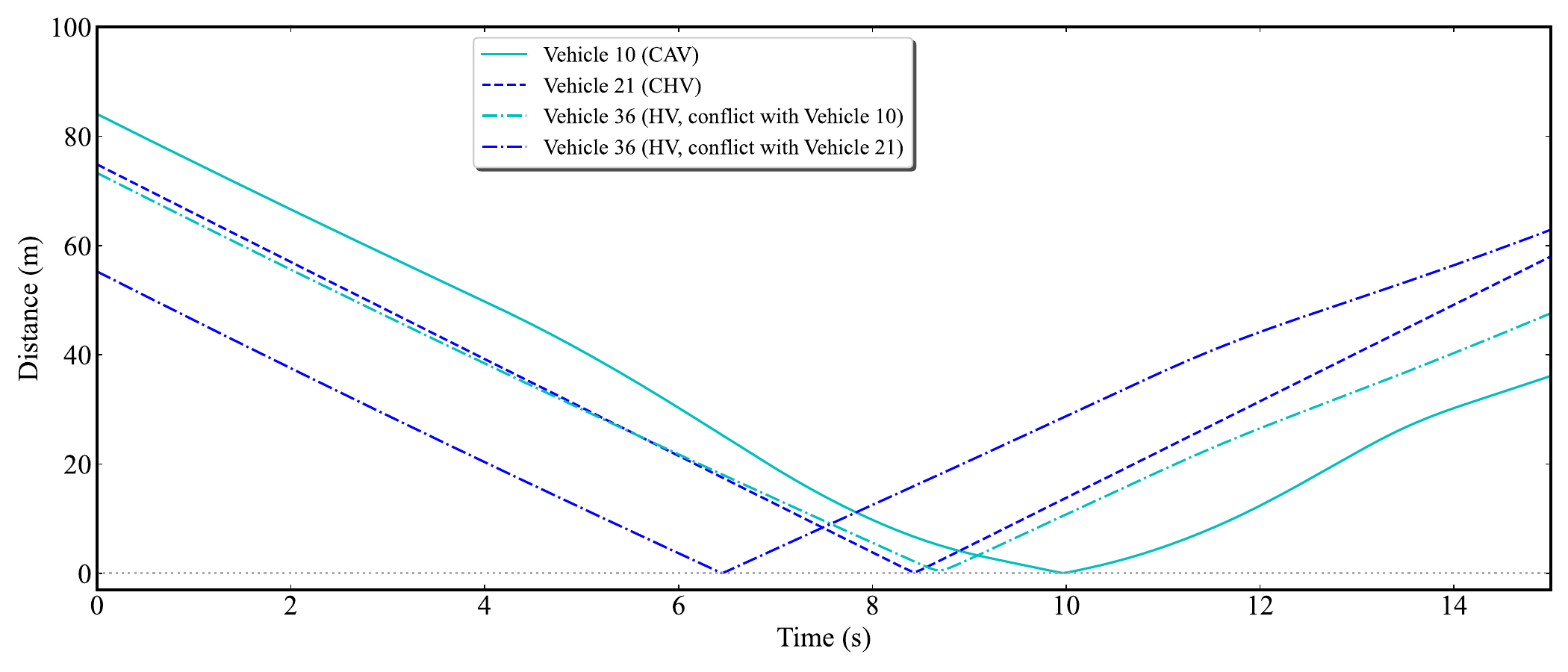}
\caption{Distance between the trajectories of the three types of vehicles to different conflict points. There are two conflict points between the three vehicles. The blue line indicates the change in distance from the vehicle to the conflict point between CAV 21 and HV 36. The cyan line indicates the change in distance from the vehicle to the conflict point between CHV 10 and HV 36.}
\label{Vehicle_position_conflict}
\end{figure}

\begin{figure}[!t]
\centering
\includegraphics[width=3.5in]{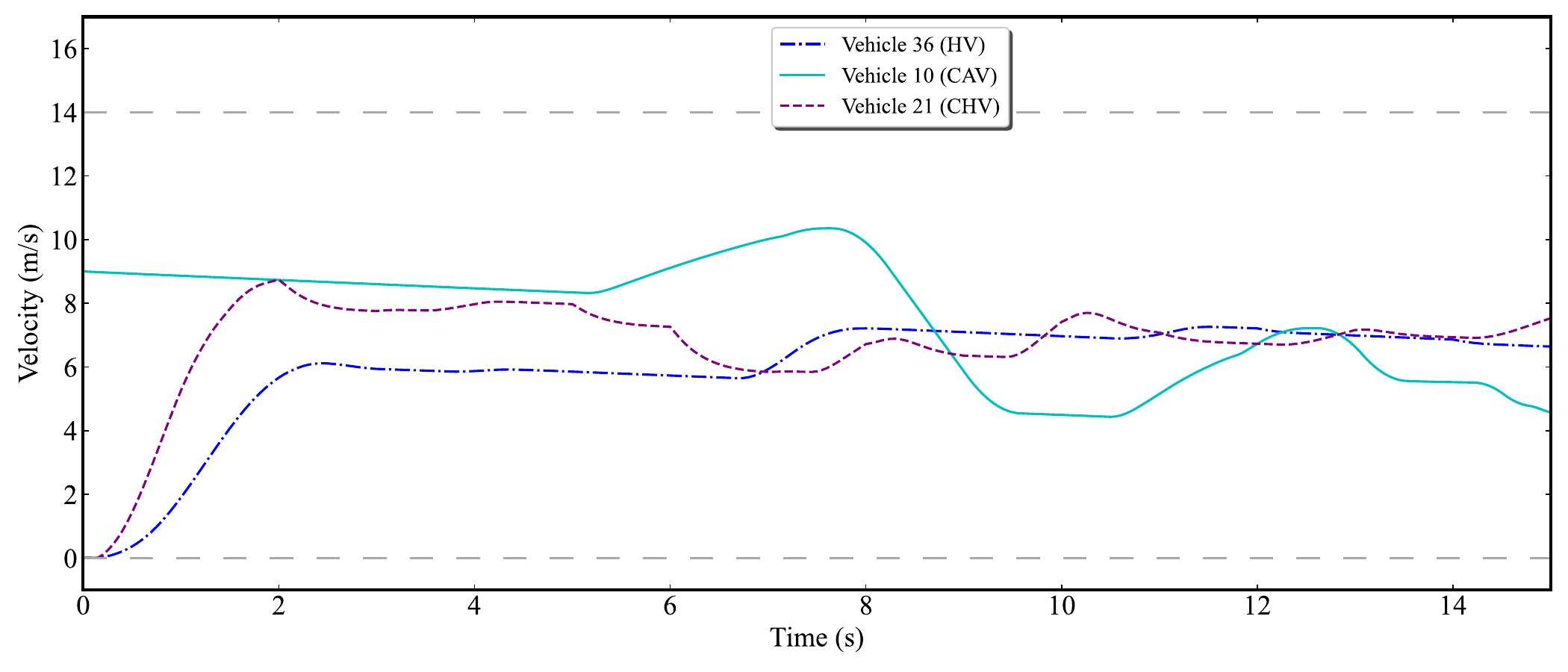}
\caption{Profiles of velocity changes for three vehicles with conflicting spatio-temporal trajectories.}
\label{fig_Vehicle_coordinate_system}
\end{figure}

Vehicle velocity profiles in the complex intersection under mixed traffic flow are illustrated in Fig. 23. As can be seen in Fig. 23, the algorithm deduces that CAVs 1-8, which are farther away from the intersection, can accelerate to the intersection in advance switching to the car following mode or conflict-solving mode to pass through the intersection quickly. Similarly, subsequent vehicles arrive at the intersection in an orderly manner under the regulation of the proposed vehicle planning algorithm. They pass through the intersection sequentially by switching to different driving modes adaptively and without disturbance according to the vehicle planning and control algorithm. 

\begin{figure}[!t]
\centering
\includegraphics[width=3.5in]{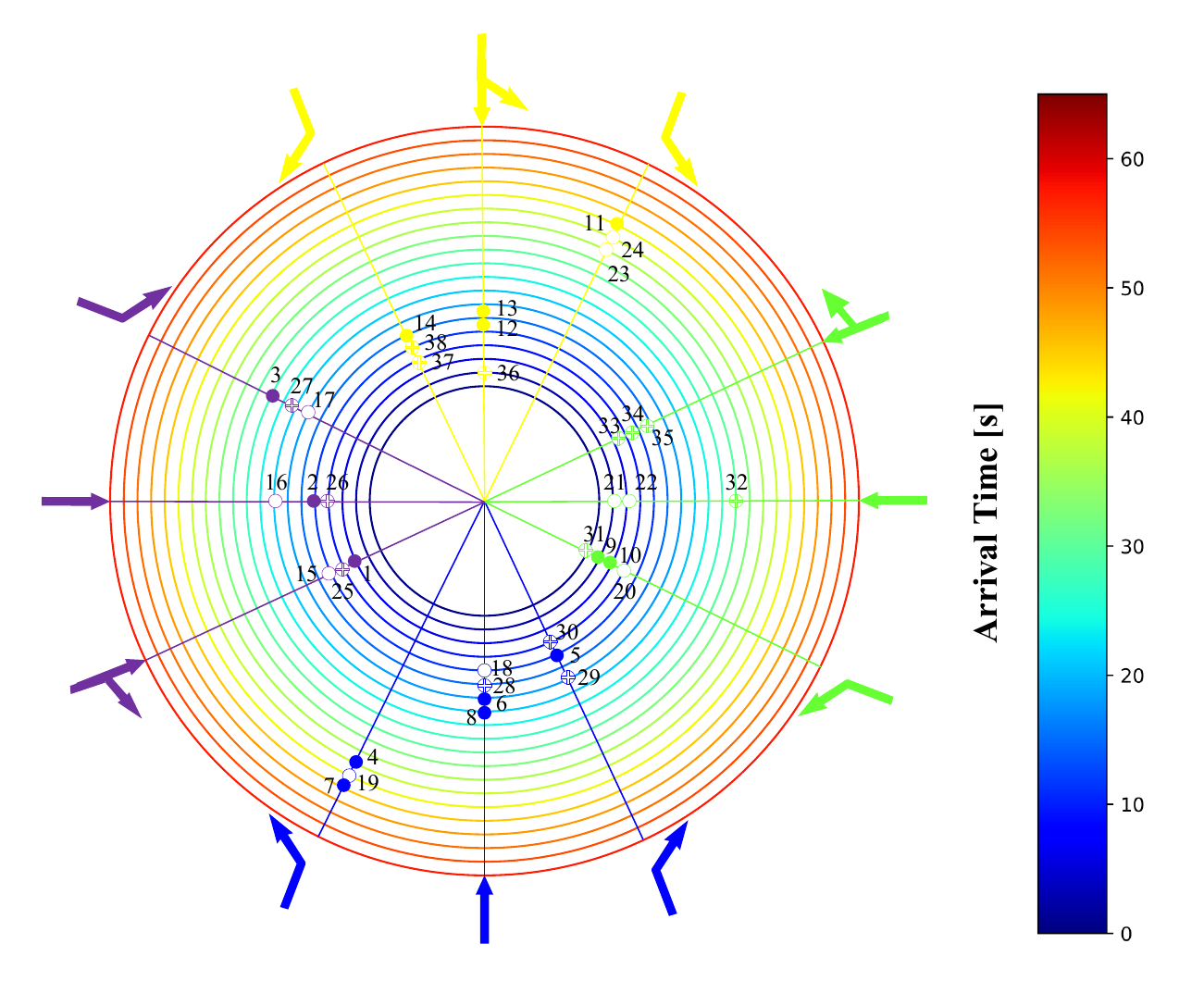}
\caption{Arrival time of vehicles from different directions at the four-way, three-lane intersection in the comprehensive simulation scenario under mixed traffic streams. The circular contours indicate different levels of arrival time, and their values are shown in the figure's color bars. The solid dots, hollow dots, and plus signs with circles represent the arrival time of CAVs, CHVs, and HVs, respectively. The dots on the same circle correspond to the vehicles that pass through the intersection at the same time.}
\label{fig_Vehicle_coordinate_system}
\end{figure}

Figs. 24–26 show the test results in the typical conflict scenario. Fig. 24 highlights the key moments of mixed traffic flow of CAVs, CHVs, and HVs by the snapshots. 
There is a typical conflict relationship between CAV 10, CHV 21, and HV 36. CAV 10 turns left through the intersection, and there is a merging conflict with HV 36, traveling straight through the intersection. CHV 21 travels straight through the intersection, and there is a crossing conflict with HV 36, which is traveling straight through the intersection. As can be seen in the figure, both CAV 10 and CHV 21 made an excellent yield to prevent the trajectory conflict with HV 36. In addition, CAV 10 utilized a conflict resolution model to follow HV 36 through the intersection as quickly as possible while ensuring safe passage.
Fig. 25 shows the distance between the trajectories of the three conflicting vehicles to different conflict points. There are two conflict points between the three vehicles. The blue line indicates the change in distance from the vehicle to the conflict point between CAV 21 and HV 36. The cyan line indicates the change in distance from the vehicle to the conflict point between CHV 10 and HV 36. All vehicles can pass through their respective conflict points in an orderly manner, maintaining a safe distance. Profiles of velocity changes for three vehicles with conflicting spatiotemporal trajectories are illustrated in Fig. 26. From the figure, it can be seen that all the conflicting vehicles can pass through the intersection smoothly at high speed, which proves the safety and efficiency of the proposed method in solving the conflict scenario with mixed traffic flow.

Fig. 27 shows the arrival time of different types of vehicles approaching the conflict zone of the intersection. The circular contours indicate different levels of arrival time, and their values are shown in the figure's color bars. The solid dots, hollow dots, and plus signs with circles in Fig. 27 represent the arrival time of CAVs, CHVs, and HVs, respectively. The dots on the same circle correspond to the vehicles that pass through the intersection at the same time. From the figure, it can be seen that there is no spatiotemporal trajectory conflict between the vehicles crossing the intersection at the same time, and all the vehicles can cross the intersection within 46.9 s, and the traffic efficiency is improved. The videos of comprehensive simulation experiments under different scenarios are uploaded online. Please visit the website to view them (\emph{https://cnc.sjtu.edu.cn/HPQ/}).

\section{CONCLUSION}
This paper proposes an intelligent intersection management system for mixed autonomy traffic streams, which comprehensively consists of algorithms for upper-level decision making and low-level vehicle planning and control. In the upper-level right of way allocation, the HPQ algorithm is designed according to the difference between CAVs, CHVs, and HVs. Since CAVs can deal with intersection trajectory conflicts with the help of intelligent equipment, they are easier to obtain the right of way than CHVs. In the low-level vehicle planning and control for CAVs, four different control modes are designed, and the control execution is carried out through the model predictive controller. The application of the model predictive controller to the CAVs can realize the undisturbed switching of the control modes and satisfy the safety requirements of collision avoidance. Then SUMO is used to verify the efficiency and superiority of the proposed HPQ algorithm, which shows that the HPQ algorithm outperforms the delay-time actuated traffic lights algorithm and the Hybrid AIM protocol by decreasing the average travel time by 5\% to 65\% for different traffic flows. Moreover, the PreScan is used to conduct a microscopic simulation at the intersection, which verifies the safety and effectiveness of the proposed vehicle planning and control algorithm. Simulation results demonstrate that the proposed method can be effectively applied to large-scale traffic intersections with mixed traffic of CAVs, CHVs, and HVs, which effectively improves the intersection efficiency under the premise of ensuring traffic safety. The proposed unsignalized intersection management strategy is practical to be applied in the near future. 

For future practical application deployment, challenges such as communication delay, packet loss, and pedestrian intervention need to be further considered. The problem will be more complicated when other traffic participants such as pedestrians are considered, because the interaction behaviors at the intersections are very diverse. We will further study the intelligent intersection management strategy in the presence of pedestrians and cyclists, so that the strategy can be applied to various types of public intersections.



%





\ifCLASSOPTIONcaptionsoff
  \newpage
\fi



\bibliographystyle{IEEEtran}
\bibliography{IEEEexample}

\end{document}